\documentstyle[psfig]{mn}

\newif\ifAMStwofonts
\def\om{{\Omega_p}}
\def\len{a_B}
\def\lag{D_L}
\def\vpd{{\cal R}}
\def\DV{{\Delta V}}
\def\pin{{\cal P}}
\def\kin{{\cal K}}
\def\smt{{\cal S}}
\def\asymk{{\rm A}_\kin}
\def\asymp{{\rm A}_\pin}
\def\kms{$\mathrm {km}~ \mathrm s^{-1}$}
\def\kmsk{$\mathrm {km}~\mathrm{s}^{-1}~ \mathrm{kpc}^{-1}$}
\def\spose#1{\hbox to 0pt{#1\hss}}
\def\gtsim{\mathrel{\spose{\lower.5ex \hbox{$\mathchar"218$}}
     \raise.4ex\hbox{$\mathchar"13E$}}}
\def\ltsim{\mathrel{\spose{\lower.5ex\hbox{$\mathchar"218$}}
     \raise.4ex\hbox{$\mathchar"13C$}}}

\def\degrees{^\circ}
\def\arcmin{^\prime}
\def\arcsec{^{\prime\prime}}
\def\ohir{OH/IR}
\def\etal{{\it et al.}}
\def\eg{{\rm e.g.}}
\def\etc{{\rm etc.}}
\def\ie{{\rm i.e.}}

\ifoldfss
  \ifCUPmtlplainloaded \else
    \NewTextAlphabet{textbfit} {cmbxti10} {}
    \NewTextAlphabet{textbfss} {cmssbx10} {}
    \NewMathAlphabet{mathbfit} {cmbxti10} {} % for math mode
    \NewMathAlphabet{mathbfss} {cmssbx10} {} %  "   "    "
  \fi
  \ifAMStwofonts
    \ifCUPmtlplainloaded \else
      \NewSymbolFont{upmath} {eurm10}
      \NewSymbolFont{AMSa} {msam10}
      \NewMathSymbol{\upi}     {0}{upmath}{19}
      \NewMathSymbol{\umu}     {0}{upmath}{16}
      \NewMathSymbol{\upartial}{0}{upmath}{40}
      \NewMathSymbol{\leqslant}{3}{AMSa}{36}
      \NewMathSymbol{\geqslant}{3}{AMSa}{3E}

      \let\leq=\leqslant \let\le=\leqslant
      \let\geq=\geqslant 
    \fi
  \fi
\fi % End of OFSS

\ifnfssone
  \newmathalphabet{\mathit}
  \addtoversion{normal}{\mathit}{cmr}{m}{it}
  \addtoversion{bold}{\mathit}{cmr}{bx}{it}
  \newmathalphabet{\mathbfit} % math mode version of \textbfit{..}
  \addtoversion{normal}{\mathbfit}{cmr}{bx}{it}
  \addtoversion{bold}{\mathbfit}{cmr}{bx}{it}
  \newmathalphabet{\mathbfss} % math mode version of \textbfss{..}
  \addtoversion{normal}{\mathbfss}{cmss}{bx}{n}
  \addtoversion{bold}{\mathbfss}{cmss}{bx}{n}
  \ifAMStwofonts
    \ifCUPmtlplainloaded \else
      \UseAMStwoboldmath
      \makeatletter
      \new@mathgroup\upmath@group
      \define@mathgroup\mv@normal\upmath@group{eur}{m}{n}
      \define@mathgroup\mv@bold\upmath@group{eur}{b}{n}
      \edef\UPM{\hexnumber\upmath@group}
      \new@mathgroup\amsa@group
      \define@mathgroup\mv@normal\amsa@group{msa}{m}{n}
      \define@mathgroup\mv@bold\amsa@group{msa}{m}{n}
      \edef\AMSa{\hexnumber\amsa@group}
      \makeatother
      \mathchardef\upi="0\UPM19
      \mathchardef\umu="0\UPM16
      \mathchardef\upartial="0\UPM40
      \mathchardef\leqslant="3\AMSa36
      \mathchardef\geqslant="3\AMSa3E

      \let\leq=\leqslant \let\le=\leqslant
      \let\geq=\geqslant 
    \fi
  \fi
\fi % End of NFSS release 1

\ifnfsstwo
  \DeclareMathAlphabet{\mathbfit}{OT1}{cmr}{bx}{it}
  \SetMathAlphabet\mathbfit{bold}{OT1}{cmr}{bx}{it}
  \DeclareMathAlphabet{\mathbfss}{OT1}{cmss}{bx}{n}
  \SetMathAlphabet\mathbfss{bold}{OT1}{cmss}{bx}{n}
  \ifAMStwofonts
    \ifCUPmtlplainloaded \else
      \DeclareSymbolFont{UPM}{U}{eur}{m}{n}
      \SetSymbolFont{UPM}{bold}{U}{eur}{b}{n}
      \DeclareSymbolFont{AMSa}{U}{msa}{m}{n}
      \DeclareMathSymbol{\upi}{0}{UPM}{"19}
      \DeclareMathSymbol{\umu}{0}{UPM}{"16}
      \DeclareMathSymbol{\upartial}{0}{UPM}{"40}
      \DeclareMathSymbol{\leqslant}{3}{AMSa}{"36}
      \DeclareMathSymbol{\geqslant}{3}{AMSa}{"3E}

      \let\leq=\leqslant \let\le=\leqslant
      \let\geq=\geqslant 
    \fi
  \fi
\fi % End of NFSS release 2

\ifCUPmtlplainloaded \else
  \ifAMStwofonts \else % If no AMS fonts
    \def\upi{\pi}
    \def\umu{\mu}
    \def\upartial{\partial}
  \fi
\fi

\title{The Pattern Speed of the \ohir\ Stars in the Milky Way}
\author[Victor P.\ Debattista, Ortwin Gerhard and Maartje N.\ Sevenster]
{Victor P.\ Debattista$^{1}$\footnotemark, 
Ortwin Gerhard$^{1}$ and Maartje N.\ Sevenster$^{2}$
\\ $^{1}$ Astronomisches Institut, Universit\"at Basel, Venusstrasse 7, 
CH-4102 Binningen, Switzerland
\\ $^{2}$ RSAA/MAAAO, RSAA/MSSSO, Cotter Road, Weston ACT 2611, 
Australia}

\date{}

\pagerange{\pageref{firstpage}--\pageref{lastpage}}
\pubyear{2001}

\begin{document}

\maketitle

\label{firstpage}

\begin{abstract}
We show how the continuity equation can be used to determine pattern 
speeds in the Milky Way Galaxy (MWG).  This method, first discussed by 
Tremaine \& Weinberg in the context of external galaxies, requires
projected positions, $(l,b)$, and line-of-sight velocities for a spatially 
complete sample of relaxed tracers.  If the local standard of rest (LSR) has 
a zero velocity in the radial direction ($u_{\rm LSR}$), then  the quantity 
that is measured is $\DV \equiv \om R_0 - V_{\rm LSR}$, where $\om$ is the 
pattern speed of the non-axisymmetric feature, $R_0$ is the distance of the 
Sun from the Galactic centre and $V_{\rm LSR}$ is the tangential motion of 
the LSR, including the circular velocity.  We use simple models to assess the 
reliability of the method for measuring a single, 
constant pattern speed of either a bar or spiral in the inner MWG.  We then 
apply the method to the \ohir\ stars in the ATCA/VLA OH 1612 MHz survey 
of Sevenster \etal, finding $\DV = 252 \pm 41$ \kms, if
$u_{\rm LSR} = 0$.  Assuming further that $R_0 = 8$ kpc and 
$V_{\rm LSR} = 220$ \kms, this gives $\om = 59\pm 5$ \kmsk\ with a possible
systematic error of perhaps 10 \kmsk.  The non-axisymmetric feature for 
which we measure this pattern speed must be in the disc of the MWG.
\end{abstract}

\begin{keywords}
Galaxy: bulge -- Galaxy: kinematics and dynamics -- 
Galaxy: disc  -- Galaxy: structure -- galaxies: spiral
\end{keywords}

\footnotetext{E-mail: debattis@astro.unibas.ch}

\section{Introduction}

Our position within the Milky Way Galaxy (MWG) has made study of its 
large-scale structure and dynamics difficult.  This is partly due to the 
Sun's position within the dust layer of the MWG, as well as the difficulty of 
determining distances.  Nevertheless, in recent years, considerable progress
has been made in understanding our Galaxy.  

It is now clear that the MWG is a barred galaxy, as was first proposed by de 
Vaucouleurs (1964).  Evidence of this comes from the {\it COBE}/DIRBE near 
infrared light distribution (Weiland \etal\ 1994; Dwek \etal\ 1995; Binney
\etal\ 1997), star count asymmetries (Nakada \etal\ 1991; Whitelock \& 
Catchpole 1992; Sevenster 1996; Nikolaev \& Weinberg 1997; Stanek \etal\ 
1997; Hammersley \etal\ 2000), gas dynamics (Peters 1975; Cohen \& Few 1976; 
Liszt \& Burton 1980; 
Gerhard \& Vietri 1986; Mulder \& Liem 1986; Binney \etal\ 1991) and the 
large microlensing optical depth towards the bulge (Paczynski \etal\ 1994; 
Zhao \etal\ 1996).  A recent review of the structure of the bulge and disc
can be found in Gerhard (2001).

The principle dynamical parameter of a bar is its pattern speed, $\om$, 
since it determines the orbital structure of stars in the disc.  Bar 
pattern speeds can be parametrized by the distance-independent ratio 
$\vpd \equiv \lag/\len$, where $\lag$ is the Lagrangian radius at which the 
gravitational and centrifugal forces cancel out in the bar rest-frame 
($\lag$ is usually approximated by the 
axisymmetric corotation radius) and $\len$ is the bar semi-major axis.  A bar 
is termed fast when $1.0 \leq \vpd \leq 1.4$.  Knowledge of bar pattern speeds 
constrains the dark matter (DM) content of disc galaxies: Debattista \& 
Sellwood (1998, 2000) argue that fast bars require that the disc they inhabit 
must be maximal, in the sense that the luminous disc provides most of the 
rotational support in the inner galaxy.  Some evidence for fast bars comes 
from hydrodynamical models of gas flow, particularly at the shocks.  Three 
such studies are: NGC 1365 ($\vpd = 1.3$, Lindblad \etal\ 1996), NGC 1300 
($\vpd = 1.3$, Lindblad \& Kristen 1996), and NGC 4123 ($\vpd = 1.2$, Weiner 
\etal\ 2001).  

In the MWG, $\om$ can be estimated by comparing hydrodynamical 
models of the inner MWG with the observed longitude-velocity, ($l,v$), 
diagram.  Several such studies have been carried out: Englmaier \& Gerhard 
(1999) found $\lag = 3.5 \pm 0.5$ kpc, $\om = 59 \pm 2$ \kmsk\ and 
$\vpd = 1.2 \pm 0.2$, while Fux (1999) found $\lag = 4 - 4.5$ kpc and 
$\om = 35 - 45$ \kmsk\ and the preferred model of Weiner \& Sellwood (1999) 
had $\lag = 5.0$ kpc, $\om = 41.9$ \kmsk\ and $\vpd = 1.4$.  Binney 
\etal\ (1991) interpreted the ($l,v$) diagram for HI, CO  and CS emission 
using orbits in a barred potential with $\om = 63$ \kmsk.  Another method 
for measuring $\om$ in the MWG involves the identification of the action 
of resonances: Binney \etal\ (1997) suggested that a local density maximum 
along the minor axis seen in the dust-corrected infrared {\it COBE}/DIRBE 
data could be identified with the location of a Lagrange point, which gave 
$\om = 65 \pm 5$ 
\kmsk, having assumed $\vpd \simeq 1.2$ and given the value of $\len$ from
the deprojection.  Dehnen (1999) argued that the bimodal distribution of 
stellar velocities in the solar neighborhood, as observed by {\it HIPPARCOS}, 
is due to the action of the outer Lindblad resonance (OLR), whereby he 
obtained $\om = 53 \pm 3$ \kmsk.  Sevenster (1999) suggested that 
features in the distribution of \ohir\ stars can be used to locate 
the corotation and inner Lindblad resonances, giving $\om \sim 60$ \kmsk\ 
($\vpd \gtsim 1.4$).  The differences in these pattern speed measurements 
reflect uncertainties in viewing geometry, $\len$, rotation curve and bar 
axis ratio.

The MWG also has spiral structure in it, which is less well 
constrained.  Even the number of arms, whether two or four, continues to be 
an issue of some discussion.  Vall\'{e}e (1995) reviews a variety of different 
observational tracers, and concludes that the spirals are four-armed, 
logarithmic and have a pitch angle, $p \simeq 12\degrees \pm 1\degrees$.  More
recently, Drimmel \& Spergel (2001) found that a two arm spiral structure
dominates the non-axisymmetric near-infrared emission.

The pattern speeds of spirals are generally quite poorly constrained.  When
a bar is present, as in the MWG, the expectation that spirals are driven by
it (\eg\ Sanders \& Huntley 1976) would seem to require that the bar and 
spirals share a common pattern speed.  This idea appears to be further 
reinforced by the observation that many such spirals connect with 
the bar ends.  Sellwood \& Sparke (1988), however, presented $N$-body
simulations in which bars and spirals had different pattern speeds, but 
with the spirals still generally connected to the bar ends.  Tagger \etal\ 
(1987) and Sygnet \etal\ (1988), proposed that the pattern speeds of the 
bar and spirals are non-linearly resonantly coupled, with the corotation of 
the bar marking the inner Lindblad resonance of the spiral.  Toomre's (1981) 
swing amplification theory of spiral structure formation requires a radially 
(and possibly temporally) varying pattern speed.  Modal theories of 
spiral formation, on the other hand, depend on a (nearly) constant spiral 
pattern speed in space (but not necessarily in time, see Bertin \& Lin 
1996).  Measurements of spiral pattern speeds in external galaxies are still 
quite few, generally assume a radially constant $\om$ and tend to give
conflicting results.  For example, the spiral pattern speed in M81 has been 
measured in a variety of ways but so far no unique pattern speed has been 
found (see discussion in Canzian 1993).  

The situation is not much better in the MWG.  Morgan (1990) was able to
model the distribution of pulsars within roughly 10 kpc of the Sun with 
spirals of $\om = 13.5$ \kmsk, regardless of the arm multiplicity.  On the 
other hand, Amaral \& L\'epine (1997), using the positions and ages of young 
open clusters within $\sim 5$ kpc of the Sun, estimated $\om \simeq 20$ 
\kmsk.  Mishurov \& Zenina (1999) used the velocity field of a sample of 
Cepheids which were less than 5 kpc from the Sun to conclude that the Sun 
is displaced outwards from the corotation by $\sim 0.1$ kpc (then $\om \simeq 
\Omega_\odot \simeq 27$ \kmsk).  

A direct method for measuring $\om$ in systems satisfying the continuity 
equation was developed by Tremaine \& Weinberg (1984).  This method has 
successfully been applied to four barred galaxies so far: NGC 936 
($\vpd = 1.4 \pm 0.3$, Merrifield \& Kuijken 1995), NGC 4596 
($\vpd = 1.15^{+0.38}_{-0.23}$, Gerssen \etal \ 1998), NGC 7079 
($\vpd = 0.9 \pm 0.15$, Debattista \& Williams 2001) and NGC 1023
($\vpd = 0.77^{+0.43}_{-0.22} $, Debattista \etal\ 2002).  In view of the
spread in previous measurements of pattern speeds in the MWG, it is worth 
considering whether this method may not also be applied it.  Kuijken \& 
Tremaine (1991) have already derived an equivalent of the Tremaine-Weinberg 
method for the MWG;
their main goal was to constrain the radial motion of the local standard of 
rest (LSR).  Applying this method to the HI distribution, they found that 
pattern speeds from $\om = 0$ to $\Omega_\odot$ are possible for an LSR 
with a radial velocity, $u_{\rm LSR} = -4$ \kms\ to 8 \kms.

In this paper, we explore the Tremaine-Weinberg method for the MWG.  This 
method, unlike others, has the considerable advantage of being model 
independent.  In the general case, the only assumption made is that the 
non-axisymmetric density distribution is in stationary cylindrical rotation 
(\ie\ $\om = \om(\rho)$, where $\rho$ is the distance from the Galactic 
centre).  While this assumption breaks down in some cases (\eg\ if the MWG 
disc is strongly interacting with a companion), it is expected to be a very 
good approximation inside the solar circle.  Since we also wish to apply 
the method to real data, we have made a number of simplifying assumptions.
For example, we assume that there is one 
unique pattern speed in the inner MWG; if this is not the case, then some 
density and asymmetry weighted average pattern speed is measured.  The value 
of $\om$ obtained depends sensitively on the assumed radial velocity of the 
LSR; however this dependence can be stated explicitly.  The method requires 
complete samples of tracers: one such sample is obtained from the
ATCA/VLA OH 1612 MHz survey of \ohir\ stars.

In \S\ref{theory}, we derive the three-dimensional version of the 
Tremaine-Weinberg method for the MWG, with the solar and LSR motion taken
into account.  We test the method in \S\ref{tests}, using simple models,
exploring under what circumstances the method works best.  In \S\ref{survey},
we introduce the ATCA/VLA OH 1612 MHz survey, and describe how we selected
a sample of older, relaxed \ohir\ stars from it.  We also discuss our 
completeness corrections for the survey data.  Then in \S\ref{results}, we 
present the pattern speed analysis for our sample of \ohir\ stars.  Since 
the integrals in the Tremaine-Weinberg method measure asymmetries between 
positive and negative longitude, they are susceptible to noise and/or
observational bias; therefore in \S\ref{results}, we present corroborating 
evidence for the signal we find.  Finally, in \S\ref{discons}, we discuss 
our results in terms of MWG structure and prospects for using the same 
method with future data sets.

\section{The Tremaine-Weinberg Method for the Milky Way}
\label{theory}

Starting from the continuity equation and the assumption that the visible 
surface density, $\mu$, of a barred (or otherwise non-axisymmetric) disc 
depends on time, $t$, in the following simple way:
\begin{equation}
\mu (x,y,t) = \mu (\rho,\psi - \om t),
\label{dens}
\end{equation}
where $(x,y)$ are arbitrary cartesian coordinates in the plane of the disc,
and $(\rho,\psi)$ the corresponding polar coordinates, Tremaine \& Weinberg 
(1984) showed that the pattern speed, $\om$, can be written as:
\begin{equation}
\om \sin i = \frac{\int^\infty_{-\infty}~ h(Y) \int^\infty_{-\infty}~ 
V_{los}(X,Y) \mu(X,Y) dX dY}{\int^\infty_{-\infty}~ h(Y) \int^\infty_{-\infty}
~ X \mu(X,Y) dX dY}.
\label{standard}
\end{equation}
Here, $V_{\rm los}$ is the line-of-sight (minus the systemic) 
velocity, $h(Y)$ is an arbitrary continuous weight function, $i$ is the 
inclination angle and $(X,Y)$ are galaxy-centered coordinates along the 
disc's apparent major and minor axis respectively.  The Tremaine-Weinberg 
(TW) method needs a tracer of the non-axisymmetric feature which satisfies 
the continuity equation.  The old stellar population in undisturbed SB0 type 
disc galaxies provides an ideal tracer of the surface density.  The short
lifetimes of massive stars precludes their use as tracers.  Moreover, gas 
cannot be used as a tracer because of its conversion between atomic and
molecular states, although modelling may be used to describe such conversions
(Bienaym\'e \etal\ 1985).

\subsection{Two dimensions}

We now seek to derive an equivalent to the TW method for the MWG.  Our 
position within the 
MWG disc gives a unique viewing geometry which does not permit the standard TW 
method of Equation (\ref{standard}) to be used.  As shown by Kuijken \& 
Tremaine (1991), however, it is still possible to use the continuity equation 
to derive an expression for a pattern speed in the MWG.  Here, we 
re-derive their two-dimensional expression before considering the
three-dimensional case.  Assuming Equation (\ref{dens}) holds, the continuity 
equation in the plane of the disc can be written:
\begin{equation}
\om \left( y {\frac {\partial \mu}{\partial x}} - x {\frac {\partial \mu}
{\partial y}} \right) + \nabla \cdot (\mu {\bf v}) = 0.
\end{equation}
Switching to Sun-centered cylindrical coordinates $(R,l)$, (where 
$l$ is the usual Galactic longitude and $R$ is the distance from the 
Sun), this becomes:
\begin{eqnarray}  
& &\om \left[ -R_0 \sin l {\frac {\partial \mu}{\partial R}} + 
\left( 1 - {\frac{R_0 \cos l}{R}} \right) {\frac {\partial \mu}{\partial 
l}} \right] \nonumber \\
& & + \frac{1}{R}{\frac {\partial (R \mu v_R)}{\partial R}}
+ \frac{1}{R} {\frac {\partial (\mu v_l)}{\partial l}} = 0.
\end{eqnarray}
In this equation, $(v_R,v_l)$ is the velocity in the $(R,l)$ frame and
$R_0$ is the distance of the Sun from the Galactic centre.  
Typically, the component $v_l$ is not available; fortunately, this term
drops out after an integration over $l$:
\begin{eqnarray}  
& & \om \left( -R_0 \frac{\partial}{\partial R} \int_{0}^{2 \pi} \mu \sin
l dl - \frac{R_0}{R}\int_{0}^{2 \pi} \mu \sin l dl 
\right) \nonumber \\
& & + \frac{1}{R}\frac{\partial}{\partial R}\int_{0}^{2 \pi}\mu v_R R 
dl = 0.
\end{eqnarray}
If we now multiply by $R g(R)$, where $g(R)$ is an arbitrary function 
vanishing at the Sun, and integrate over $R$, we obtain, after some 
rearrangement:
\begin{equation}
\om R_0 = \frac{\int_0^\infty \int_0^{2 \pi} \mu f(R) v_R \ 
dl~ dR}{\int_0^\infty \int_0^{2 \pi} \mu f(R) \sin l \ 
dl~ dR},
\label{int2d}
\end{equation}
where we have assumed that $\mu g(R)$ goes to zero at infinity and $f(R) 
\equiv R \partial g/\partial R$.  All the quantities on the right 
hand side of the above equation are observable.

In the case of discrete tracers, the surface density gets replaced by a sum 
over delta functions and the above expression is replaced by:
\begin{equation}
\om R_0 = \frac{ \sum_{i} f(R_i) v_{R,i}}{\sum_{i} f(R_i) \sin l_i}
\label{inert2d}.
\end{equation}
Thus $f(R)$ can be thought of as a detection probability for the
discrete tracers.

In Equation (\ref{int2d}) above, we 
integrated $l$ from $0$ to $2 \pi$.  But if the non-axisymmetric part of 
$\mu$ is limited to the range $[-l_0,l_0]$, it is then trivial to 
show that these integration limits are sufficient, since 
axisymmetric components do not contribute to the integrals, and all surface 
terms resulting from the $l$ integrals vanish by symmetry.  It is necessary,
however, that the integrals extend from $-l_0$ to $l_0$ (\ie\ integration 
in the range $[-l_1,l_2]$, with $l_1 \neq l_2$ gives incorrect results.)

When the bar is either along or perpendicular to the line of sight, 
the numerator and denominator of Equation (\ref{inert2d}) are zero, and 
$\om R_0$ cannot be determined.  This is to be compared with the standard TW
equation, where $\om$ cannot be measured for a bar along the apparent major 
or minor axis of the disc.

\subsection{Three dimensions}

In 3D, the derivation is more or less the same.  The Sun-centered spherical
polar coordinates are now $(r,l,b)$, with $(l,b)$ being the standard
galactic coordinates and $r$ the distance from the Sun.  Following a 
triple integral much in the spirit of the 2D case, we obtain:
\begin{equation}
\om R_0 = \frac{\int_0^\infty \int_{-b_0}^{+b_0} \int_{-l_0}^{+l_0} 
f(r)~ \mu~ \cos b~ v_r~ dl~ db~ dr} {\int_0^\infty \int_{-b_0}^{+b_0} 
\int_{-l_0}^{+l_0} f(r)~ \mu~ \cos^2 b~ \sin l~  dl~ db~ dr},
\label{threed}
\end{equation}
where $b_0$ must be such that all the non-axisymmetric parts 
are included.  If the system is symmetric about the mid-plane, then the 
$b$ integral can extend over the region $[0,\pm b_0]$ only: data of this kind 
are easier to obtain.  A discrete version of this equation is given by:
\begin{equation}
\om R_0 = {\frac{\sum_{i} f(r_i) v_{r,i}}{\sum_{i} f(r_i) \sin l_i \cos b_i}}
\equiv \frac{\kin(l_0)}{\pin(l_0)}.
\label{inert3d}
\end{equation}
where we have introduced a new notation for the integrals.  (Note that one 
factor of $\cos b$ drops out because of the Jacobian in the delta functions.)  
In the limit that all the tracers are in a plane, 
Equation (\ref{inert3d}) reduces to Equation (\ref{inert2d}) of the 2D case.

Both $\pin$ and $\kin$ are defined as integrals/sums over $l$ ranging from
$-l_0$ to $l_0$.  Since $\sin l$ and $v_r$ change sign across 
$l = 0$, the integrals will partially cancel out.  Thus $\pin$ and $\kin$
largely measure a difference across $l = 0$.  When the difference is 
small, $\pin$ and $\kin$ are small and the measured $\om R_0$ will be poorly
determined.  This can be seen from the expression for the variance in 
$\om R_0$: $\left(\frac{\sigma_{[\om R_0]}}{\om R_0}\right)^2 = 
 \left(\frac{\sigma_\kin}{\kin}\right)^2 + 
 \left(\frac{\sigma_\pin}{\pin}\right)^2$.
To be more explicit by what we mean by small, we define
$\pin_\pm \equiv \int_0^\infty \int_{-b_0}^{+b_0} \int_{0}^{\pm l_0} f(r) 
\mu \cos^2 b \sin l \ dl~ db~ dr$, with a similar definition for 
$\kin_\pm$.  (In this notation $\pin = \pin_+ - \pin_-$ and similarly 
for $\kin$.)  Then we introduce the variables
\begin{equation}
\asymk = \frac{|\kin|}{|\kin_-| + |\kin_+|}, \ \ \ 
\asymp = \frac{|\pin|}{|\pin_-| + |\pin_+|}.
\end{equation}
These two parameters, which take values in the range $[0,1]$, quantify the 
degree of global asymmetry in the system.  When $\asymp = \asymk = 0$, no 
asymmetry signal is present (but note that the system need not be 
axisymmetric, or even symmetric in projection) while when 
$\asymp = \asymk = 1$ the observed system is maximally asymmetric.  In
\S\ref{tests}, we use simple models to explore the dependence of measurement 
errors on $\asymp$ and $\asymk$.

\subsection{More complicated density distributions}
\label{complications}

Several complications may arise, including multiple pattern speeds, 
non-stationary pattern speeds, solar motion, \etc\  Here we discuss issues 
related to violations of the density assumption in Eqn. 
\ref{dens}, deferring discussion of solar motion to 
\S\ref{subsec:solar_motion}.  Possible violations of Eqn. \ref{dens} include:

\begin{enumerate}

\item {\it Multiple pattern speeds.}  The MWG has a bar and spirals, perhaps 
also a small lopsidedness; almost certainly all have different pattern 
speeds.  To understand how these will affect the measurement, we suppose that 
there are $n$ non-axisymmetric features superposed, so that Eqn. \ref{dens} 
becomes $\mu (x,y,t) = \sum_{i=0}^n\mu_i (\rho,\psi - \Omega_i t)$.  Each of 
these asymmetries gives rise to integrals $\pin_i$ and $\kin_i$.  Then, by the 
linearity of the continuity equation, we can write 
$R_0 \sum_i \Omega_i \pin_i = \sum_i \kin_i \equiv \kin$.  Dividing both 
sides by $\pin \equiv \sum_i \pin_i $, we obtain 
$R_0 \overline{\Omega} = \kin/\pin$, where $\overline{\Omega}$ is the
average of the different pattern speeds, weighted by the $\pin_i$'s.  Thus
the average is weighted by both the asymmetry and the density (since if 
$\mu_1 = s \mu_2$ then $\pin_1 = s \pin_2$).  Therefore, 
small perturbations to the disk density do not contribute very significantly 
to $\overline{\Omega}$.  However, since the $\pin_i$'s can be either positive 
or negative, $\overline{\Omega}$ is not limited to values in the range 
$\min(\Omega_i)$ and $\max(\Omega_i)$ (unless all the $\pin_i$'s have the 
same sign).  If distances are available, $f(R)$ can be varied, to test
for such cancellations; otherwise, strong changes in the slope of $\pin$ as 
a function of $l_0$ may be used to test for cancellations.

Note that winding spiral arms are a special case of the multiple pattern
speed situation.  In this case, $\om$ is a continuously varying function of
distance from the Galactic center.  This case is discussed further in
\S\ref{future}.

\item {\it Radial oscillations.}  If such oscillations were present, 
Eqn. \ref{dens} would become $\mu (x,y,t) = \mu (\rho + 
\rho_1 e^{-i \omega(t-t_0)},\psi - \om t)$, where we assume $\rho_1$ is
small.  The time-derivative term in the continuity equation therefore becomes
\begin{equation}
\frac{\partial \mu}{\partial t} \simeq 
-\om \frac{\partial \mu}{\partial \psi} 
-i\omega \rho_1 e^{i \omega t_0} \frac{\partial \mu}{\partial R}
\equiv A_0 + A_1.
\end{equation} 
The ratio $|A_1/A_0|$ is then given by $\omega \rho_1/(2 \om R_d)$, where,
for concreteness, we have assumed that the disk is exponential with 
scale-length $R_d$ and the azimuthal density variation is of the form 
$\mu(R) e^{2 i \psi}$.  For a simple estimate, we assume 
$\omega \sim \max(\kappa) \sim 2\Omega$.  Therefore we obtain 
$|A_1/A_0| \sim V_c \rho_1/\om R_d \rho$; at $\rho = 1$ kpc, if the circular 
velocity is 220 \kms, $\om = 60$ \kmsk\ and $R_d = 2.1$ kpc (Bissantz 
\etal\ 2002), we obtain $|A_1/A_0| \sim 1.7 \rho_1$ kpc$^{-1}$.  Thus radial 
oscillations with amplitude larger than about 100 pc will strongly 
interfere with the measurement of pattern speeds using the TW method.
However, radial oscillations are expected to be strongly Landau damped, and 
it would be very surprising if the MWG exhibits large radial oscillations.  

\item {\it Growing amplitudes.}  In this case, we can replace 
Eqn. \ref{dens} with $\mu (x,y,t) = \mu_0 (\rho) + \mu_1 (\rho,\psi - \om t) 
e^{\alpha t}$, where $\alpha > 0$ ($\alpha < 0$) represents a growing 
(damped) feature.  The term $\mu_0 (\rho)$ is included to make explicit
the fact that the density of the feature grows at the expense of a 
background axisymmetric disc (for which $\pin = \kin = 0$), to conserve
the total mass.  The time-derivative term is now replaced by
\begin{equation}
\frac{\partial \mu}{\partial t} = 
-\om e^{\alpha t} \frac{\partial \mu}{\partial \psi} 
+\alpha \mu_1 e^{\alpha t}
\equiv A_0 + A_1.
\end{equation}
It is possible to propagate the term in $\alpha \mu_1$ through the 
calculation, giving, in the 2D case
\begin{equation}
\om R_0 \pin + \alpha \int_0^\infty \int_0^{2 \pi} R g(R) \mu_1 \ dl~ dR = \kin.
\label{growing}
\end{equation}
Eqn. \ref{growing} contains both $f(R)$ and $g(R)$, and therefore requires
distance data for a solution.  Lacking distances, we estimate 
the interference from growing features by the ratio $|A_1/A_0|$, which,
for an $m=2$ azimuthal density variation, is given by $|\alpha|/(2 \om)$.
Thus the growth term is important when the growth rate is of order 
the pattern speed (in which case it is safe to ignore damping).  Since
rapid growth is mostly restricted to the low amplitude, linear regime,
while $\asymp$ can only be large enough for an accurate measurement once 
growth is nearly saturated, we expect that rapid growth is unlikely to
significantly interfere with pattern speed measurements.

The most important growing structures are spirals.  
If they are modal, they can grow even when tightly wound.  The 
theoretically computed growth rates vary substantially; \eg\ the 
fastest growing two arm modes in the flat-rotation, cut-out discs 
of Evans \& Read (1998) have $0.03 \ltsim \alpha/2\om \ltsim 0.29$.  

\item {\it Time-varying pattern speeds.}  In this case, Eqn. \ref{dens} 
gets replaced by $\mu (x,y,t) = \mu (\rho,\psi - \om(t) t)$.  Although
this introduces a new term, proportional to 
$t \frac{\partial \om}{\partial t}$, it also changes the angle derivative 
to one in the bar rest frame.  A simple transformation to a time 
$t^\prime$ in which the moment of observation is $t^\prime = 0$ recovers
Eqn. \ref{threed}.  Thus a time-varying pattern speed does not 
disturb the TW method.  

\end{enumerate}

We have considered several violations of the density condition of Eqn. 
\ref{dens}.  Of these, perhaps the most likely are growing spirals, 
but it is unclear how rapidly growing spirals are.  Most likely, if 
spirals are of sufficient amplitude to give large signals, then their
growth rate cannot be too large.  This may be sorted with more detailed data 
on the MWG density distribution than we currently have, such as will be 
available with future astrometric missions.
For the present work, we have assumed that there are no (strong) 
non-axisymmetric structures with rapidly growing amplitudes.

\subsection{Solar motion}
\label{subsec:solar_motion}

We derived Equation (\ref{threed}) for an inertial frame.  Any radial 
velocities that can be measured will be heliocentric, and therefore 
the motion of the Sun needs to be taken into account. The Sun's velocity can 
be decomposed into two parts: a motion of the local standard of rest, 
$\bf{V_{LSR}}$, and the peculiar motion of the Sun relative to the LSR, 
$\bf{v_{\odot,pec}}$.  We assume that the motion of the LSR consists of
two parts, a tangential motion around the MWG centre with velocity 
${\bf v_{LSR}} = u_{\rm LSR}{\bf \hat{x}} + V_{\rm LSR}{\bf \hat{y}}$, 
where $\bf{\hat{x}}$ is in the direction towards the Galactic centre and 
$\bf{\hat{y}}$ is the in-plane tangential direction.  Then, if velocities 
in the heliocentric frame are labelled as $\bf{v}_i^\prime$, we can 
write $v_{r,i} = 
({\bf{v_i}^\prime} + {\bf{v}_{\odot}})\cdot {\bf\hat{r}_i} = 
v_{r,i}^\prime + {\bf{v}_{\odot,pec}}\cdot{\bf\hat{r}_i} + 
{\bf{v}_{LSR}}\cdot {\bf\hat{r}_i} = v_{r,i}^\prime + 
{\bf{v}_{\odot,pec}}\cdot{\bf\hat{r}_i} + V_{\rm LSR} \sin l_i \cos b_i
+ u_{LSR} \cos l_i \cos b_i$.  Thus Equation (\ref{inert3d}) can be rewritten:
\begin{eqnarray}  
\DV & \equiv & \om R_0 - V_{\rm LSR} \nonumber \\
    & = & \frac{ \sum_{i} f(r_i)(v_{r,i}^\prime + 
{\bf{v}_{\odot,pec}}\cdot{\bf\hat{r}}_i) }{\sum_{i} f(r_i)\sin l_i 
\cos b_i} \nonumber \\
& & - u_{LSR} 
\frac{ \sum_{i} f(r_i)\cos l_i \cos b_i}{\sum_{i} f(r_i)\sin l_i \cos b_i} 
\nonumber \\
& \equiv & \frac{\kin}{\pin} - u_{\rm LSR} \frac{\smt}{\pin}.
\label{sunmotion}
\end{eqnarray}
This equation, which is the 3D form of the TW equation derived by 
Kuijken and Tremaine (1991), is used in the analysis of the \ohir\ star 
sample below.  It contains two quantities, $V_{\rm LSR}$ and $u_{\rm LSR}$ 
which must be determined from other data.  Here, $V_{\rm LSR}$ includes 
the circular velocity; Reid \etal\ (1999) measured $V_{\rm LSR} = 219 \pm 20$ 
\kms\ for $R_0 = 8$ kpc, from the apparent proper motion of Sgr A$^*$, 
while from a similar measurement Backer \& Sramek measured 
$V_{\rm LSR} = 234 \pm 7$.  The uncertainty 
in $u_{\rm LSR}$ leads to a much larger uncertainty in $\DV$, largely
because $\smt$ depends on the $\cos l_i$'s, which do not cancel across $l=0$.  
The TW equation for external galaxies is similarly sensitive to the relative
motion of the observer and the observed galaxy, but in that case reflection
symmetry is used to minimize this sensitivity.  In the absence of distance
information, no similar symmetry argument can be deployed in the MWG.  
The solar peculiar motion, ${\bf{v}_{\odot,pec}}$, has 
been measured with {\it HIPPARCOS} data by Dehnen \& Binney (1998).  They 
found ${\bf{v}_{\odot,pec}} = (U_0,V_0,W_0) = (10.00,5.25,7.17) \pm 
(0.36,0.62,0.38)$ \kms.  (These correspond to velocity 
components that are radially inwards, in the direction of Galactic rotation 
and vertically upwards, respectively.)  In the absence of a radial motion of
the LSR, we will then be measuring $\DV \equiv \om R_0 - V_{\rm LSR}$, \ie\ 
the negative of the solar velocity in the bar's rest frame.

In the coordinate system we are working in, a star moving away from the LSR
has a positive $v_{r}$; inside the solar circle, such stars will primarily be
at positive $l$.  Therefore, if a positive $\DV$ is measured, this will
correspond to a non-axisymmetric feature in the inner MWG with a pattern
speed $\om > \Omega_\odot$, which must therefore be rotating in the same
sense as the disc.

The variance in $\DV$ due to the errors in the Sun's measured 
peculiar velocity is given by:
\begin{eqnarray}
\sigma^2_{\DV} & = &
 \sigma^2_{U_0} \frac{\left( \sum_i f(r_i) \cos l_i \cos b_i\right)^2}
{\pin^2} \ + \ \sigma^2_{V_0} \nonumber \\
& &+  \sigma^2_{W_0} \frac{\left( \sum_i f(r_i) \sin b_i\right)^2}{\pin^2}
\label{errsolpecvel}
\end{eqnarray}
(where we have assumed that errors in position measurements are 
insignificant).  Note that, in this expression, the dominant term is the 
coefficient of $\sigma^2_{U_0}$, since $\cos l$ is everywhere positive in 
the range $-\frac{\pi}{2} \leq l \leq \frac{\pi}{2}$.

\section{Tests}
\label{tests}

We now test the TW method on some simple models.  Our goal is not so much
to test for specific structures expected in the MWG, since such an approach 
would say more about our models, which we would like to keep simple.
Rather, our goal here is to learn how reliable the TW method is under
various {\it observational} constraints, such as the number of discrete 
tracers, signal strength, restricted ranges of $l$, \etc\  Errors from
sources specific to our application are discussed in \S\ref{results}.

Models used to test the TW method need to have velocity fields which 
preserve the assumed density distribution, \ie\ the assumption of Eqn. 
(\ref{dens}) needs to remain valid.  However, because the continuity equation 
is purely kinematic, a model density distribution need not generate the 
potential required by the kinematics.  Here we test the method on
models for a bar and a spiral, with application to the \ohir\ stars in
the ATCA/VLA OH 1612 MHz survey of Sevenster \etal\ (1997a,b \& 2001) in 
mind.  Since \ohir\ stars are concentrated in the inner MWG, our tests 
assume that the non-axisymmetric structure is inside the Solar circle.  We 
assume throughout that measurements of $l$, $b$ and $v_r$ for individual 
stars have only small errors, which we ignore, as is appropriate for this 
survey.

\subsection{Bar pattern speeds}

\begin{figure}
\begin{center}
\leavevmode\psfig{figure=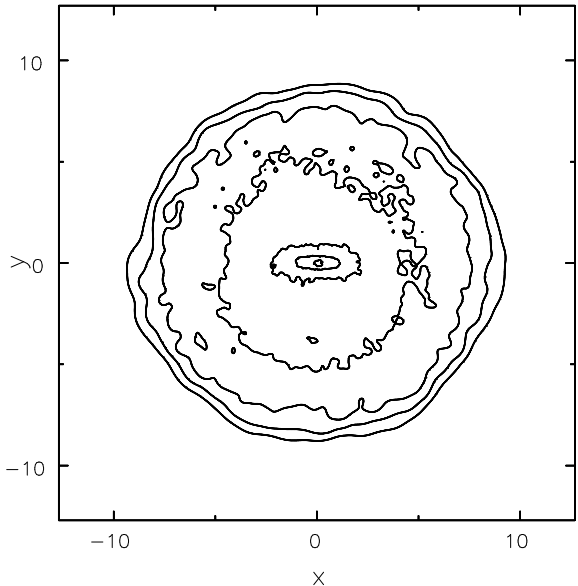,width=0.9\hsize}
\end{center}
\caption{Contours of projected disc density of the $N$-body barred system 
used in the tests.  Contours are logarithmically spaced.  This is the
``maximum disk'' simulation of Debattista \& Sellwood (2000) at $t=650$.}
\label{fig01}
\end{figure}

Tests of the method for a realistic bar can be obtained from 
$N$-body simulations.  We tested Equation (\ref{inert3d}) for a selection 
of the bars in the fully self-consistent, $N$-body simulations of Debattista 
\& Sellwood (2000), who measured $\om$ from the time rate of change of the
phase of the $m=2$ Fourier component of the disc density.  Here we present 
the results of several experiments using their ``maximum disk'' simulation 
at $t=650$, which is shown in Fig. \ref{fig01}, when the bar was fast 
($\vpd = 1.1 \pm 0.2$).  At this time, $\len = (2.6 \pm 0.6) R_d$ (where 
$R_d$ is the disc scale-length), which is about twice the size of the MWG 
bar.  The viewer therefore was placed at $4 \leq R_0/R_d \leq 5$ (within the
disc).  The angle of the bar to the 
Sun-center line, $\psi$, was varied in the expected range 
($20\degrees \leq \psi \leq 30\degrees$).  The angle $l_0$, within which 
the integrals $\pin$ and 
$\kin$ were evaluated, was varied in the range $30\degrees \leq l_0 \leq 
60\degrees$.  The simulation included $102,000$ disc particles; here we
describe two experiments in which samples of 500 particles were selected.  In 
the first experiment, particles were drawn using $f(r)$ a Gaussian with
$\overline{r} = 2.25R_d$ and $\sigma_r = 0.7R_d$.  The second experiment used 
a cosine probability: $f(r) = \cos(\frac{r}{R_{max}}\frac{\pi}{2})$ with 
$R_{max} = 6R_d$ for $r \le R_{max}$ and $f(r) = 0$ otherwise.  

\begin{figure*}
\begin{center}
\leavevmode{\psfig{figure=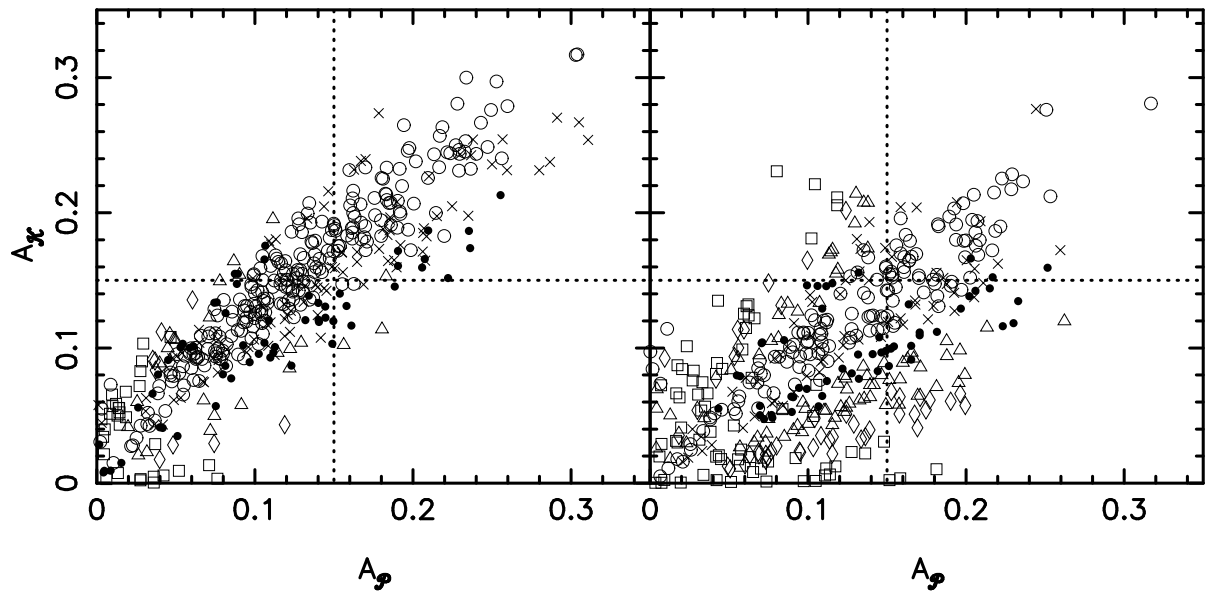,width=0.9\hsize,clip=,angle=0}}
\end{center}
\caption{The fractional errors in the measurement of $\DV$ for a fast bar
using 500 particles.  The left panel is for the Gaussian probability and
the right panel for the cosine probability.  The symbols represent 
(absolute) errors as follows: circles 
$0.0 \leq |\delta(\DV)/\DV| < 0.2$, crosses $0.2 \leq |\delta(\DV)/\DV| < 
0.3$, filled circles $0.3 \leq |\delta(\DV)/\DV| < 0.4$, triangles $0.4 \leq 
|\delta(\DV)/\DV| < 0.6$, diamonds $0.6 \leq |\delta(\DV)/\DV| < 0.8$ and
squares $0.8 \leq |\delta(\DV)/\DV|$.}
\label{fig03}
\end{figure*}

Fig. \ref{fig03} plots the results, from which it is clear that the best 
results are obtained at large $\asymp$ and $\asymk$; since the Gaussian 
probability selects particles in a more limited radial range, it often gives 
larger $\asymp$ and $\asymk$ and smaller errors.  However, the cosine 
probability is likely to be more realistic for a flux-density-limited 
sample of tracers, since it preferentially selects nearby objects over
distant ones.  Fig. \ref{fig03} show that, when 
$\asymp \gtsim 0.15$ and $\asymk \gtsim 0.15$, the expected error is 
on average $\sim 17\%$, and never worse than $40\%$.

Variations in $\psi$, $l_0$ and $R_0$ mattered to the errors in $\DV$ only 
insofar as they changed $\asymp$ and $\asymk$.  For the Gaussian $f(r)$
experiment, increasing $R_0$ decreased $\asymp$ and $\asymk$, with a 
corresponding increase in the fractional error.  However, changes in $\psi$ 
(within our range) and 
$l_0$ did not substantially change $\asymp$ and $\asymk$, and the error is
not overly sensitive to variations in these parameters.  For the cosine
probability distribution, $\asymp$ and $\asymk$ were insensitive to 
variations in all 3 parameters.  It is particularly reassuring that the
errors do not depend on $l_0$, provided that the bar is inside $l_0$.

\subsection{Spiral pattern speeds}

\begin{figure}
\begin{center}
\leavevmode{\psfig{figure=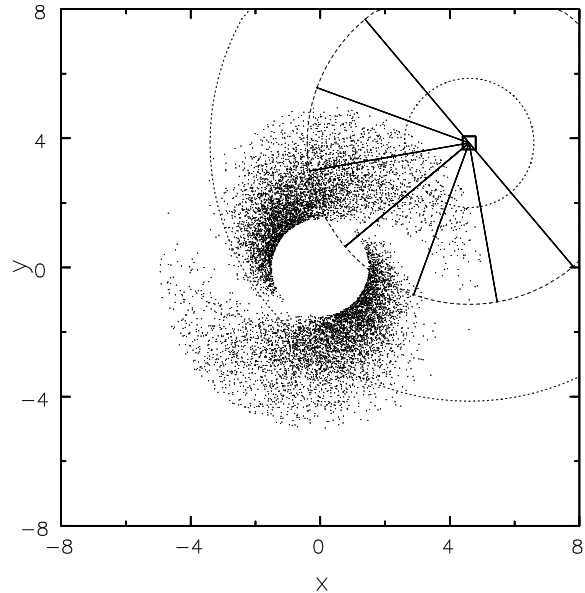,width=0.9\hsize,clip=,angle=0}}
\end{center}
\caption{The simple spiral system.  The 10,000 particles shown 
are all from the spiral component; for clarity, the axisymmetric component 
has not been plotted.  The observer, for the orientation of
Fig. \ref{fig05}, is marked by a square, with the line segments indicating 
$l = 0\degrees,\ \pm 30\degrees,\ \pm 60 \degrees$ and $\pm 90 \degrees$.  The 
dashed circle then has a radius $\overline{r} = 5R_d$ and the dotted circles 
have radius $\overline{r} \pm \sigma_r = 5 \pm 3R_d$.}
\label{fig04}
\end{figure}

Stationary spiral models based on density wave theory are somewhat harder
to construct.  Since we have already demonstrated that the TW method works
well for bars, which have larger non-circular motions than do spirals,
a crude spiral model with purely circular orbits suffices for a spiral
test.  We therefore
generated an exponential disc in the radial range 
$1.5R_d \leq r \leq 5R_d$, consisting of two components, an axisymmetric 
component with particles on circular orbits at velocity 2.5 in some 
arbitrary units and a material (rather than wave) spiral part, which 
consisted of particles on circular orbits with constant $\om = 1$.  The 
spiral density varied tangentially as $1+\cos 2 (\theta-\theta_0)$, where 
$\theta_0 = \om t - \ln(r) \cot\gamma$, to produce two-armed logarithmic 
spirals of pitch-angle $\gamma$, which we set to $20\degrees$.  Both 
components were modelled by a Gaussian vertically, with equal 
scale-heights.  The axisymmetric component accounted for $\frac{2}{3}$ of 
the total mass, producing a spiral contrast consistent with that observed 
in external galaxies (Rix \& Zaritsky 1995).  The spiral component is 
shown in Fig. \ref{fig04}.

Once we generated this system, we computed the integrals
of Equation \ref{inert3d} for a variety of orientations.  
In Fig. \ref{fig06}, we plot the errors at  $l_0 \geq 30\degrees$ as before 
for the bar; in all cases the error is $\ltsim 20\%$ when 
$\asymp \gtsim 0.15$ and $\asymk \gtsim 0.15$.
Since our rather crude spiral model included no random motions, we expect
that more realistic errors will resemble those we found for the bar, 
provided signals are large.  

\begin{figure*}
\begin{center}
\leavevmode{\psfig{figure=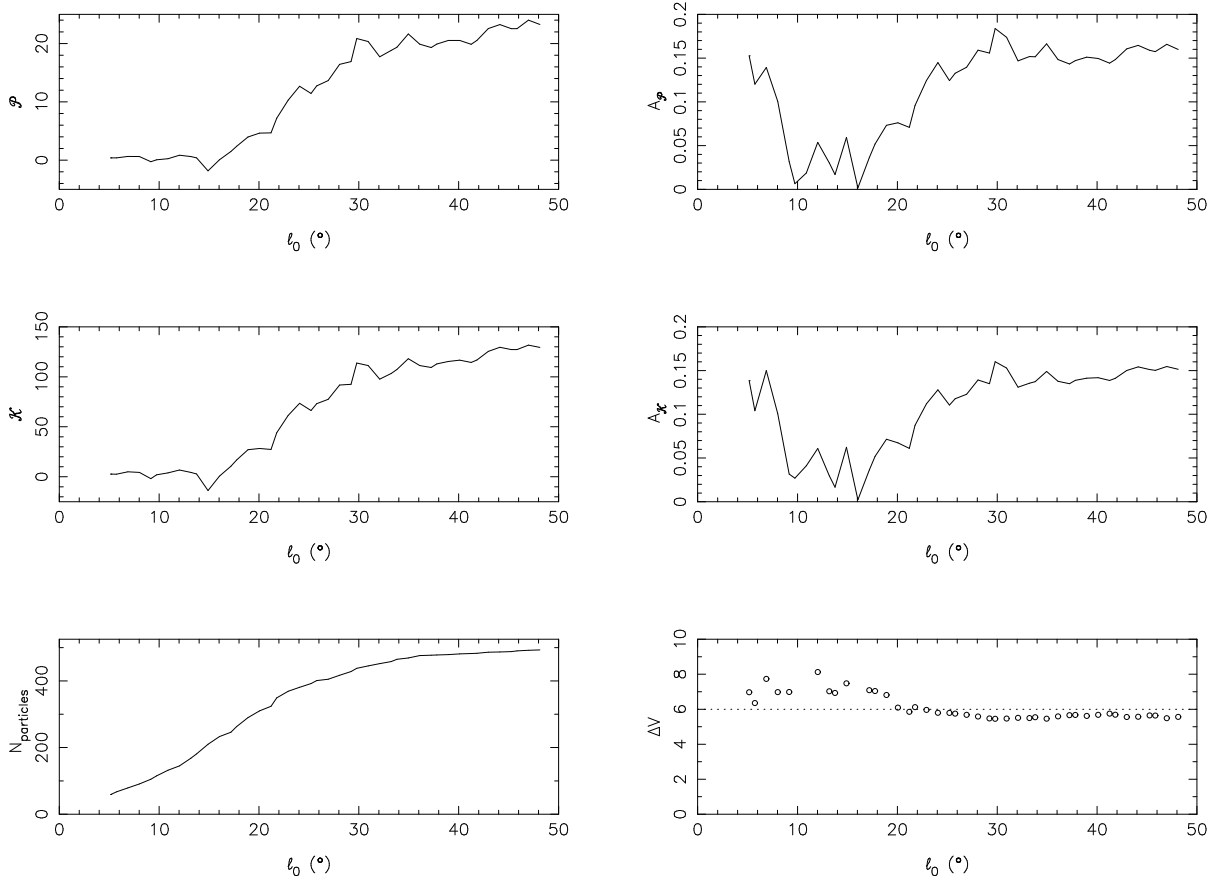,width=0.9\hsize,clip=,angle=-90}}
\end{center}
\caption{The results of analysis with 500 particles drawn from the simple
spiral model in the orientation of Fig. \ref{fig04}.  In the top 
row are $\pin$ and $\asymp$, in the middle row 
$\kin$ and $\asymk$, and in the bottom row is shown the number of particles, 
N$_{\rm particles}$, and $\DV$ obtained.  The correct value of
$\DV$ is shown by the dotted line.}
\label{fig05}
\end{figure*}

\begin{figure}
\begin{center}
\leavevmode{\psfig{figure=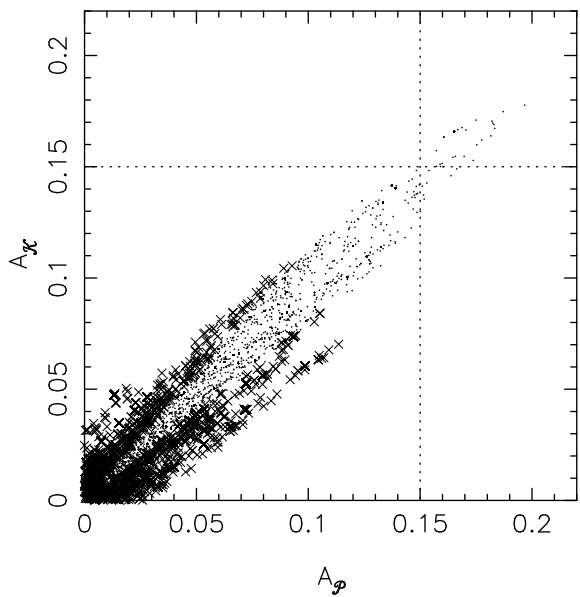,width=0.9\hsize,clip=,angle=0}}
\end{center}
\caption{Errors in $\DV$ in the analyses of simple spiral models.  The
sub-samples include different radial probability functions, orientations and 
$30\degrees \leq l_0 \leq 45\degrees$.  Points have 
$|\delta(\DV)/\DV| \leq 20\%$ while crosses have 
$|\delta(\DV)/\DV| > 20\%$.  Clearly, in all cases where $\asymp \gtsim 0.15$ 
and $\asymk \gtsim 0.15$, the expected error in $\DV$ is $\ltsim 20\%$.}
\label{fig06}
\end{figure}

Our analysis of one orientation on a sample of 500 stars, selected with 
a Gaussian probability of $\overline{r} = 5R_d$ and $\sigma_r = 3R_d$, is 
shown in Fig. \ref{fig05}; at $l_0 = 45\degrees$, the error on $\DV$ is $7.5\%$
for this case.  Note that $\asymp$ and $\asymk$ are both less than 0.2 at
$l_0 = 45\degrees$, which is smaller than the signal seen in the \ohir\
stars (\S\ref{results}).  The results for this orientation bear a striking
resemblance to those of the \ohir\ stars discussed in \S\ref{apattspd}.

We ran a variety of additional tests of the 
TW method on our simple spiral system, with varying system orientation, 
radial probability function, number of particles selected, and reduced 
spiral mass fraction.  These cases gave results consistent with those shown 
in Fig. \ref{fig06}; unsurprisingly, we found that the error decreases as 
the number of particles is increased.

\subsection{Synthesis of the tests}

We have shown that, for 500 stars, we expect to be able to measure $\DV$ 
with average errors of $\sim 17\%$, and always less than $40\%$, for either 
a bar or a spiral, provided that signals are large ($\asymp$ and $\asymk$ 
$\gtsim 0.15$).  Furthermore, we found that limiting $l_0$ does not lead 
to large errors, provided that the full non-axisymmetric structure is 
included.  

In addition to the tests described above, we also experimented with other 
parameters, with results generally in agreement with those presented
here.  The notable exception was the case of discontinuous $f(r)$'s, which
gave large errors.  Such errors can be understood in terms of noisy surface
terms in the integrals.  Lacking distance information, $f(r)$ can only be
a continuous function, corresponding to the detection probability.

\section{The \ohir\ Star Catalogue}
\label{survey}
\subsection{\ohir\ stars}

\ohir\ stars are oxygen-rich, cool AGB stars in the superwind phase (Renzini
1981) with typical $\dot{M} \sim 10^{-5}$ M$_{\odot}$yr$^{-1}$.  The wind
outflow velocity, $V_{\rm e}$, is $\sim 10 - 30$ \kms, with $V_{\rm e}$ 
larger in the
more massive (younger) \ohir\ stars.  The mass-loss rate is large 
compared to the stellar mass (typically, 1 to 6 M$_\odot$) and the superwind 
phase is therefore believed to be very short-lived $\sim 10^{5-6}$ yr 
(Whitelock \& Feast 1993).  Thus they are rare objects.  On the other hand, 
they are old objects, with ages $\sim 1-8$ Gyr, and are therefore dynamically 
relatively relaxed.  The dusty circumstellar envelope which forms from the 
outflowing material absorbs the 
stellar radiation and re-emits it in the infrared,
pumping OH masers (Elitzur \etal\ 1976) in the process. In
optically-thick envelopes, the strongest
line of this maser is at 1612.23 MHz which is conveniently insensitive
to interstellar extinction.  The Doppler shifted profile 
from the front and back of the thin OH shell 
permits easy identification of \ohir\ stars, which makes possible an unbiased 
survey of these objects, tracing the kinematics of the inner MWG.  
A full review of the properties of \ohir\ stars can be found in Habing (1996).

\subsection{The ATCA/VLA OH 1612 MHz survey}

We have used the data from the \ohir\ star catalogue to search for pattern 
speeds in the inner MWG.  The data were obtained in 3 surveys: the ATCA-bulge 
and ATCA-disc surveys (Sevenster \etal\ 1997a,b) at the Australia Telescope 
Compact Array (ATCA), covering $-45.25 \degrees \leq l \leq 10.25 \degrees$, 
$|b| \leq 3.25\degrees$ and the VLA-disc survey (Sevenster \etal\ 2001) at 
the Very Large Array (VLA), covering 
$ 4.75 \degrees \leq l \leq 45.25 \degrees$, $|b| \leq 3.25
\degrees$.  The ATCA region was covered uniformly with 1449 pointings.  On 
the other hand, only $92\%$ of the VLA region was surveyed, with 965 pointings 
out of 1053 giving useful data.  In all, the surveys produced 793 detections.  
The ATCA-bulge and VLA-disc surveys have a small overlap region 
($4.75\degrees \leq l \leq 10.25\degrees$).  In this overlap region, 27 
sources were observed in both surveys, thus only 766 independent sources 
were found.  However, the overlap region also included detections in either 
only the VLA-disc or only the ATCA-bulge survey, which is, in part, a 
manifestation of the intrinsic variability of \ohir\ stars.

Details of the reduction technique for the 3 surveys are contained in
Sevenster \etal\ (1997a,b) and Sevenster \etal\ (2001).  Of the 793 detections,
no outflow velocity was measured in 105 sources.  Since such objects can be
supergiants or even star-forming regions, rather than \ohir\ stars, and
don't have well--determined velocities, we have excluded them from our 
analysis.  

\begin{figure}
\begin{center}
\leavevmode{\psfig{figure=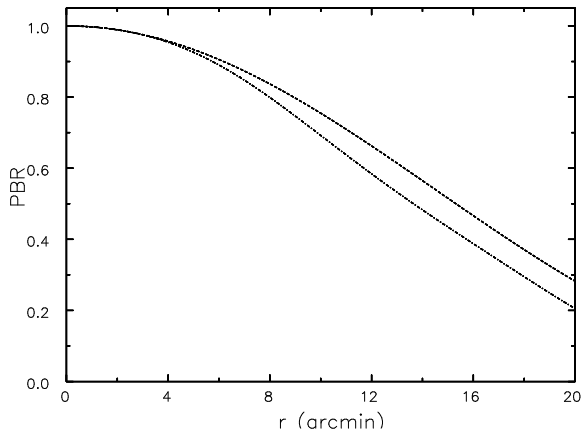,width=0.9\hsize,clip=,angle=-90}}
\end{center}
\caption{The PBR for the two instruments used in the \ohir\ survey.  The 
dashed (dot-dashed) line is the PBR for the ATCA (VLA).}
\label{fig07}
\end{figure}

The use of two separate instruments for compiling the catalogue introduces 
a systematic difference in the completenesses of the two sides of the MWG.  
We corrected for the (empirically determined) completenesses, rewriting
Equation (\ref{threed}):
\begin{equation}
\DV = \frac{\sum_{p}\left(C_p^{-1}\sum_{s\in p} v_{r,s} \right)}
{\sum_{p}\left(C_p^{-1}\sum_{s\in p} \sin l_s \cos b_s \right)},
\label{ohirtw}
\end{equation}
where $\sum_{p}$ is a sum over pointings, $\sum_{s\in p}$ is a sum over stars 
detected in pointing $p$ and $C_p$ is the completeness of pointing $p$.  The
completenesses consist of two terms: $C_p = C_{p,a} C_{p,n}$.  The first term, 
$C_{p,a}$, is a measure of the fraction of the surveyed area that is within 
some threshold radius.  For this purpose, all VLA pointings, including the
missing ones, were treated identically, and the $C_{p,a}$'s were calculated by
Monte-Carlo integration.  The threshold radii were set by requiring a constant 
primary beam response (PBR) threshold, equal in both telescopes, for each 
pointing centre.  The PBR of the two instruments is shown in
Fig. \ref{fig07}; Table \ref{tbcpa} lists several PBR levels, with the
corresponding offset radii and values of $C_{p,a}$.

\begin{table*}
\centering
\begin{minipage}{140mm}
\caption{The VLA and ATCA maximum offsets and the resulting values of 
$C_{p,a}$ for various PBR levels.}
\label{tbcpa}
\begin{tabular}{ccccc} \\
PBR level & 
VLA offset ($\arcmin$) &
$C_{p,a}$ (VLA) & 
ATCA offset ($\arcmin$) &
$C_{p,a}$ (ATCA) \\[15pt]
 0.5 & 13.6 & 0.656 & 15.3 & 0.812 \\
 0.6 & 11.7 & 0.489 & 13.3 & 0.623 \\
 0.7 &  9.9 & 0.347 & 11.2 & 0.450 \\
\end{tabular}
\end{minipage}
\end{table*}

\begin{figure}
\begin{center}
\leavevmode{\psfig{figure=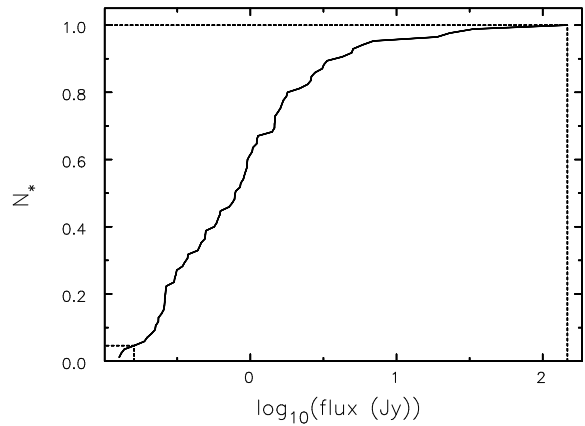,width=0.9\hsize,clip=,angle=-90}}
\end{center}
\caption{The cumulative flux distribution of \ohir\ stars in the
ATCA/VLA OH 1612 MHz survey.  We used this flux 
distribution, defined from all sources with highest peak observed at PBR 
$> 0.8$, to compute the noise completenesses, $C_{p,n}$ of the pointings.  
The lower limit is at 0.16 Jy while the upper limit is here set at the 
brightest observed \ohir\ star (147.6 Jy).}
\label{fig08}
\end{figure}

\begin{figure}
\begin{center}
\leavevmode{\psfig{figure=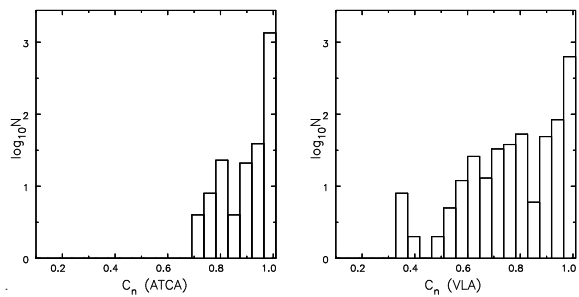,width=0.9\hsize,clip=,angle=0}}
\end{center}
\caption{The distribution of the $C_{p,n}$'s.  On the left are shown the
ATCA pointings, and on the right are the VLA pointings.  The effect of the
higher noise level in the VLA-disc survey is easily apparent.}
\label{fig09}
\end{figure}

The second term, $C_{p,n}$, is a completeness due to the noise level, and 
varies from pointing to pointing.  The $C_{p,n}$'s were obtained from the 
observed \ohir\ star cumulative flux density distribution, $N_*$, shown 
in Fig. \ref{fig08}, as the fraction of stars between the flux limits which 
are brighter than the detection threshold of each pointing.  This is computed 
in the following way:
\begin{equation}
C_{p,n} = \frac{N_*(\max(f_{min},N_s \sigma_p) \leq f \leq f_{max})}
                 {N_*(f_{min} \leq f \leq f_{max})}.
\end{equation}
Here, $f_{min}$ and $f_{max}$ are the flux density limits we use, 
$\sigma_p$ is the noise in pointing $p$ and $N_s$ is the multiplicative
detection threshold ($N_s = 7,\ 4,\ 6$ empirically determined 
for VLA-disc, ATCA-bulge, ATCA-disc respectively).  We chose 
$f_{min} = 0.16$ Jy, since this is the highest of the lower absolute detection 
thresholds of the 3 surveys, and no upper limit on $f$.  The noise level 
was more variable in the VLA-disc survey then in the ATCA surveys.  In 
Fig. \ref{fig09}, we plot the resulting distribution of the $C_{p,n}$'s.

The three surveys had different velocity coverages: the ATCA-bulge survey
covered $-280$ \kms\ $\leq v_r \leq 300$ \kms, the ATCA-disc survey covered
$-295$ \kms\ $\leq v_r \leq 379$ \kms, while the VLA-disc survey had full
coverage in the range $-200$ \kms\ $\leq v_r \leq 210$ \kms.  Thus the
three surveys had full {\it common} velocity coverage from -200 \kms\ to 200 
\kms.  A considerable number of detections were obtained at $|v_r| > 200$ 
\kms, but these are mostly at $|l|\leq 5\degrees$.  One detection
was obtained at 270 \kms\ in the ATCA-disc survey, but as this turned out 
to be a single-peaked object anyway, it was excluded from our analysis.

Typical errors in positions are $\leq 1\arcsec$ for the ATCA surveys and 
$2\arcsec$ for VLA-disc survey.  These position errors introduce negligible
errors in the value of $\DV$, as can be appreciated from Equation 
\ref{threed}.  The velocity errors are of order 1 \kms\ for 
the ATCA surveys, and 2.5 \kms\ for the VLA-disc survey.  The velocity errors 
produce larger errors in $\DV$, although, as will be seen below, these
are still quite small.

The values of $v_r$ reported in Sevenster \etal\
(1997a,b and 2001) included a correction for an assumed solar peculiar 
motion of 19.7 \kms\ towards RA = 18:07:50.3, Dec = +30:00:52 
(J2000.0).  For this study, we are adopting the solar peculiar motion found
by Dehnen \& Binney (1998); we have therefore had to transform from the 
Sevenster \etal\ frame to that of Dehnen \& Binney.

The mass, and therefore age and luminosity, of \ohir\ stars correlate with 
$V_{\rm e}$.  For example, Sevenster (2001) finds that \ohir\ stars with 
$V_{\rm e} = 13\ (17.5)$ \kms\ have a mass of roughly $1.7\ (4)M_\odot$ and 
therefore an age of 1.8\ (0.2) Gyr.  We can anticipate, therefore, that we
will need to make cuts on $V_{\rm e}$.  However, the $V_{\rm e}$'s have 
been measured in units of half the channel-widths (2.27 \kms\ for the 
VLA-disc survey, and 1.46 \kms\ for the ATCA surveys).  To compensate for 
this effective binning, we have randomized the outflow velocities within 
each bin.  We did this in a number of ways with similar results once random
re-samplings were performed; we have therefore relied on the simplest 
approach, namely to add to each $V_{\rm e}$ a random value distributed 
uniformly between -1.2 \kms\ and 1.2 \kms.

\begin{figure}
\begin{center}
\leavevmode{\psfig{figure=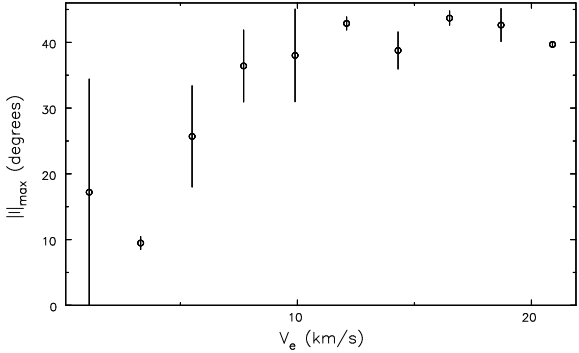,width=0.9\hsize,clip=,angle=0}}
\end{center}
\caption{The dependence of $|l|_{\rm max}$ on $V_{\rm e}$, showing that only
the nearest of the low $V_{\rm e}$ \ohir\ stars are visible in the survey.
From this figure, we conservatively estimate that only at $V_{\rm e} \geq 10$ 
\kms\ are stars visible throughout the survey region ($|l| \leq 45
\degrees$).  Each point is the average of the positive and negative $l$, 
while the error bars are defined as half the difference.}
\label{fig10}
\end{figure}

The luminosity, $L_*$, of \ohir\ stars increases with $V_{\rm e}$: simple
models predict $V_{\rm e}^{4} \propto L_*$, while more sophisticated 
models predict $V_{\rm e}^{3.3} \propto L_*$ (Habing \etal\ 1994).  Thus a 
flux density-limited sample will probe to different distances for different 
$V_{\rm e}$.  In Fig. \ref{fig10} we plot the maximum longitude, 
$|l|_{\rm max}$, of \ohir\ stars versus $V_{\rm e}$, from which it can be 
seen that \ohir\ stars of low $V_{\rm e}$ are found only at small $|l|$.  We 
interpret this as resulting from the ATCA/VLA OH 1612 MHz survey being able 
to detect only the very nearest 
faint \ohir\ stars of a distribution ending inside the solar circle.  For the
faint stars, the effect of small scale number density fluctuations 
is large, both because only a small portion of the disc is seen and because 
the number of such stars is low.  We have therefore chosen to exclude all 
\ohir\ stars with $V_{\rm e} < 10$ \kms.  

The TW method requires a sample of stars that form a relatively relaxed
population; a convenient cutoff criterion would be at least 2.5 rotations 
at 6 kpc.  This requires ages greater than 0.5 Gyr, or 
$V_{\rm e} \leq 16$ \kms.  However, since we have had to randomize the 
outflow velocities we used only stars with $V_{\rm e} \leq 15$ \kms, to 
minimize contamination by very young stars.  Thus our final range in outflow 
velocities is $10$ \kms\ $\leq V_{\rm e} \leq 15$ \kms, which is expected to 
cover a distance range of 4 to 10 kpc from the Sun.  The lower limit removes 
roughly $10\%$ of stars from the sample, while the upper limit removes a 
further $35\%$.

\section{Results}
\label{results}
\subsection{A pattern speed in the disc}
\label{apattspd}
\begin{figure*}
\begin{center}
\leavevmode{\psfig{figure=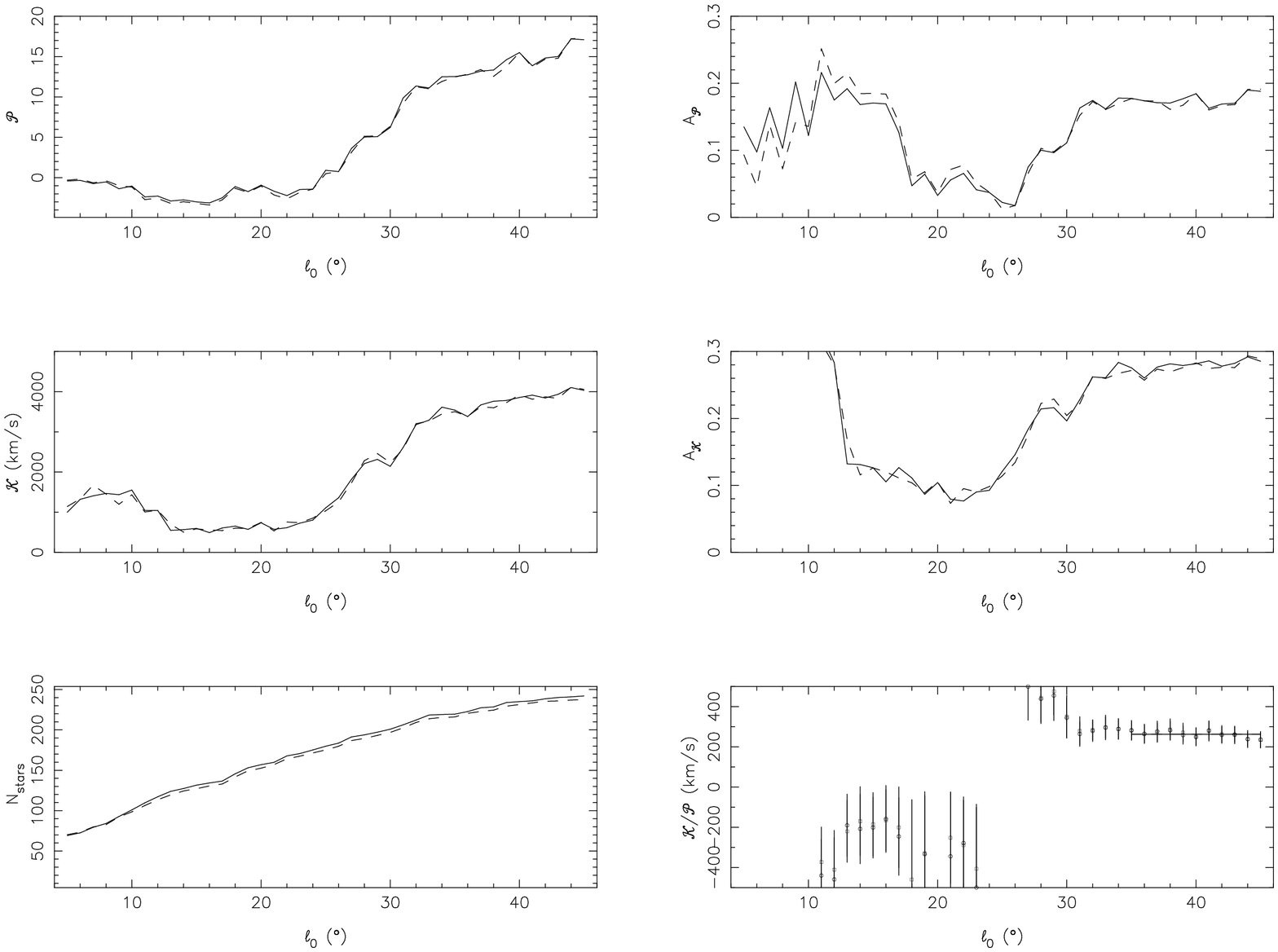,width=0.9\hsize,clip=,angle=-90}}
\end{center}
\caption{The TW analysis for the \ohir\ stars as functions of $l_0$.  In 
the top row are $\pin$ and $\asymp$, in the middle row 
$\kin$ and $\asymk$, and in the bottom row is shown the number of stars, 
N$_{\rm stars}$, and the resulting $\kin/\pin$.  In all panels, the solid
line corresponds to using the ATCA survey in the overlap region, while 
the dashed line shows the result of using the VLA survey.  In the 
bottom-right panel, circles (squares) are for the ATCA (VLA) survey, and 
the horizontal lines indicate the value of $\kin/\pin$ obtained by averaging 
results at $l_0 \geq 35\degrees$.  The values of $\kin$ and the ($1\sigma$) 
error bars on $\kin/\pin$ were obtained from 100 Monte-Carlo iterations as 
described in the text.}
\label{fig11}
\end{figure*}

We computed $\pin$, $\kin$, $\asymp$ and $\asymk$ for $5\degrees \leq l_0
\leq 45\degrees$ in steps of $1\degrees$.  We set $f_{min} = 0.16$ Jy, 
PBR = 0.5 and excluded \ohir\ stars with $|v_r| > 280$ 
\kms.  We tried both using the ATCA and the VLA in the overlap 
region.  At each $l_0$, the values reflect averages over 100 Monte-Carlo 
experiments containing slightly different samples.  In these experiments, 
the $V_{\rm e}$'s were randomized in their bins (as described above) and 
Gaussian random errors of $\sigma_v = 5$ \kms\ added to the radial 
velocities.  These results are presented in Fig. \ref{fig11}.  It is 
immediately clear that there is effectively no difference between using
the ATCA or the VLA in the overlap region.  Therefore, since the ATCA survey 
tends to be less 
noisy, we use it from here on.  For $l_0 \leq 25\degrees$, both $\pin$ and
$\kin$ are small and roughly constant, while $\asymp$ and $\asymk$ are noisy,
suggesting they are strongly affected by Poisson noise.  In the range 
$25\degrees \leq l_0 \leq 33\degrees$, $\pin$, $\kin$, $\asymp$ and $\asymk$
all grow very rapidly.  At $l_0 \simeq 33\degrees$, there is a sudden change
in the slope of the number of included stars, which is associated with almost
constant $\pin$, $\kin$, $\asymp$ and $\asymk$ at $l_0 \geq 33\degrees$.
In this region, both $\asymp$ and $\asymk$ are $\gtsim 0.15$, which our simple
tests of \S\ref{tests} suggest is large enough for moderate error levels.  
Averaging over $l_0 \geq 35\degrees$, we obtain $\kin/\pin = 263 \pm 44$ \kms, 
where we have used one half of the difference between the maximum and minimum 
value for the error estimate since the values of $\kin/\pin$ at different 
$l_0$ are not independent.  Then, if $u_{\rm LSR} = 0$, $R_0= 8$ kpc and 
$V_{\rm LSR} = 220$ \kms\ (which we assume for the remainder of this paper 
except where otherwise noted), $\om = 60 \pm 6$ \kmsk.

\begin{figure}
\begin{center}
\leavevmode{\psfig{figure=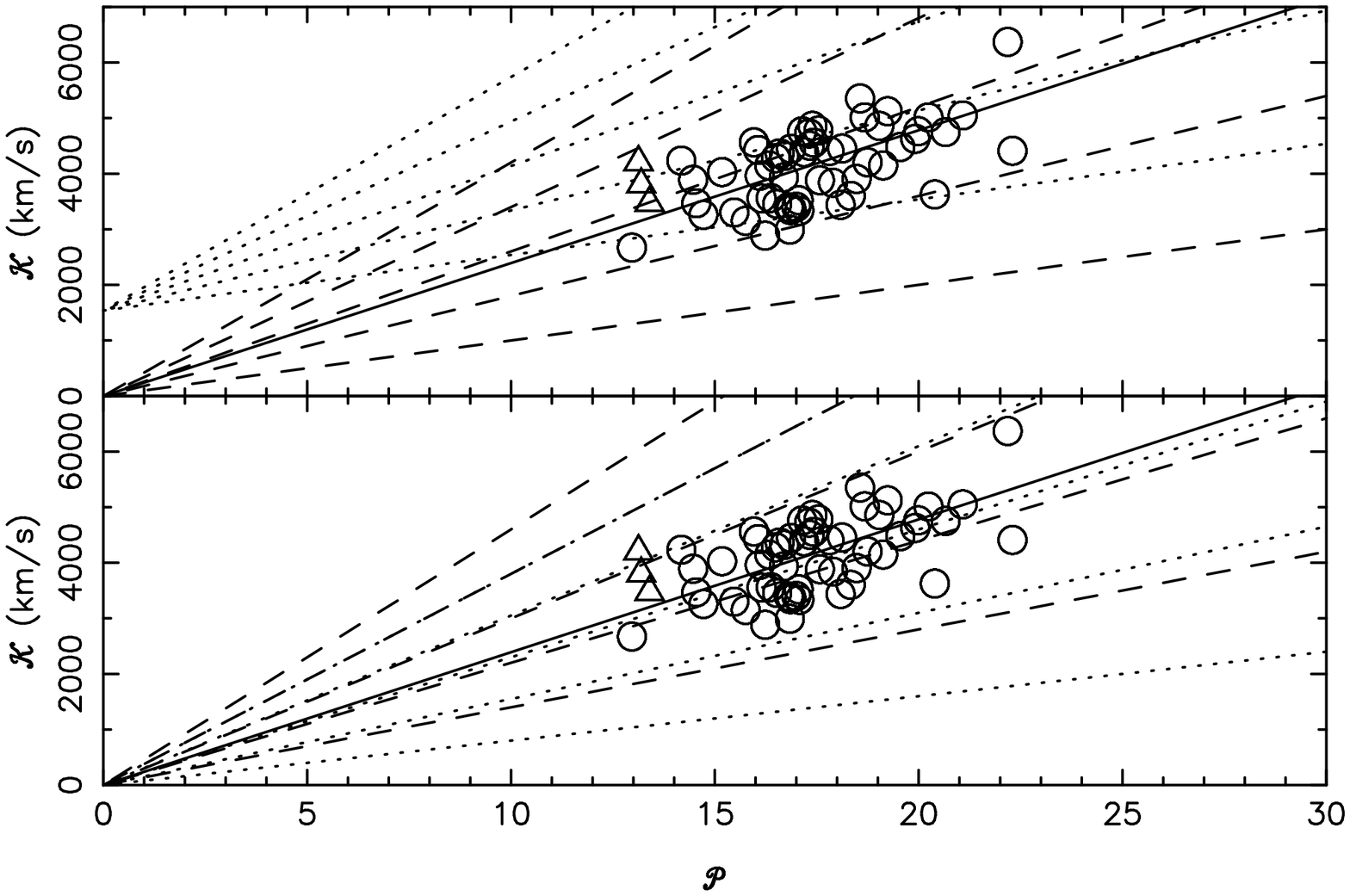,width=0.9\hsize,clip=,angle=-90}}
\end{center}
\caption{The values of $\kin$ and $\pin$ at $l_0=45\degrees$ with bounds
$10 \leq V_{\rm e} \leq 15$ \kms\ after reshuffling the $V_{\rm e}$'s.  Each 
point represents a different reshuffling.  Circles have $\asymp \geq 0.15$ and
$\asymk \geq 0.15$, while triangles have either $\asymp$ or $\asymk$ or both 
less than 0.15.  In the top panel, the dashed lines and dotted lines have 
$\om =$ 40, 50, 60, 70 and 80 \kmsk\ (in order of increasing slope) assuming 
$R_0 = 8$ kpc and $V_{\rm LSR} = 220$ \kms.  The dashed lines assume 
$u_{\rm LSR} = 0$, while $u_{\rm LSR} = 5$ \kms\ for the dotted lines.  The 
solid line (in both panesls) is the average $\om = 57$ \kmsk.  In the bottom 
panel, we show the effect of changing $V_{\rm LSR}$ to 180 \kms\ (dashed
lines) and $R_0$ to 7.5 kpc (dotted lines).}
\label{fig12}
\end{figure}

Another way of presenting the data of Fig. \ref{fig11} is shown in Fig. 
\ref{fig12}, where we plot the same data for $l_0 = 45\degrees$ but without the 
averaging of the various values of $\kin$ and $\pin$ resulting from the 
Monte-Carlo reshufflings of $V_{\rm e}$.  In the upper panel of this figure, 
we also plot lines of constant $\om$ (40, 50, 60, 70 and 80 \kmsk) assuming 
$R_0 = 8$ kpc and $V_{\rm LSR} = 220$ \kms\ for $u_{\rm LSR} = 0$ and for 
$u_{\rm LSR} = 5$ \kms.  For the $u_{\rm LSR} = 0$ case, the various 
re-samplings have average $\om = 57$ \kmsk, while in the case of 
$u_{\rm LSR} = 5$ \kms, average $\om = 45$ \kmsk, graphically demonstrating 
the sensitivity of the value of $\om$ obtained to a small radial LSR motion.  
The bottom panel also shows the effect of varying $V_{\rm LSR}$ and $R_0$, 
demonstrating that $\om$ increases with decreasing $R_0$ and/or increasing 
$V_{\rm LSR}$.  However, we will not be concerned here with errors on 
$V_{\rm LSR}$, $R_0$ and $u_{\rm LSR}$; $V_{\rm LSR}$ enters only in the
ratio $V_{\rm LSR}/R_0$, which has been determined to $10\%$ accuracy, $R_0$
is determined to even higher accuracy, while the effect of $u_{\rm LSR}$ is
considered in \S\ref{lsrradmot}.  Note that the value of $\DV$ obtained this 
way is slightly different from that obtained by averaging over $l_0$; since 
the values of $\DV$ at different $l_0$ are not independent, we quote values 
of $\DV$ and $\om$ based on $l_0 = 45\degrees$ only from now on.

\subsection{Is it real?}

\begin{figure}
\begin{center}
\leavevmode{\psfig{figure=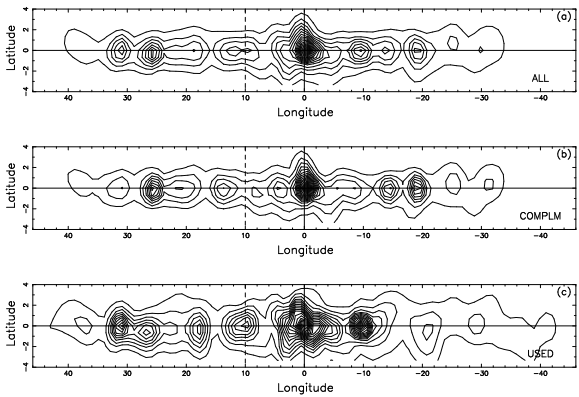,width=0.9\hsize,clip=,angle=-90}}
\end{center}
\caption{The surface density of \ohir\ stars detected in the surveys.  
Adaptive smoothing has been used.  The
surface density has not been corrected for differences in the detection
efficiencies between the ATCA ($-45\degrees \leq l \leq 10.25\degrees$)
and the VLA ($10.25\degrees \leq l \leq 45\degrees$).  Therefore the two 
sides of the plot (separated by the vertical dashed line) need to be 
considered separately.  Note the very 
strong over-densities at $25\degrees \ltsim l \ltsim 32\degrees$).  The 
bottom panel shows the sample of \ohir\ stars used in our analysis.}
\label{fig13}
\end{figure}

Before proceeding further, we present evidence that the 
signal we have measured is real, and not a result of a systematic error in 
either our analysis or the \ohir\ star surveys.  Since $\kin$ and $\pin$
measure differences, any systematic error in our calculated 
completenesses of the VLA or ATCA survey will be amplified into a signal.  
Several reasons lead us to believe that this has not happened.  Firstly, 
the signals of Fig. \ref{fig11} first become evident at a very clear
region in the disc, which is well correlated with an observed structure.  
In Fig. \ref{fig13}, we present contour maps of the surface density of
\ohir\ stars in the survey.  These clearly show the presence of very
significant over-densities localized at $l \simeq 25\degrees$ and 
$l \simeq 32\degrees$.
While, there are roughly equal number of stars at positive and negative 
$l$, ({\it despite the missing pointings, smaller} PBHW {\it and greater 
noise in the} VLA survey), for $25\degrees < |l| < 33\degrees$, there are 13 
\ohir\ stars at negative $l$ and 24 at positive $l$, which has less than 
$10\%$ probability, even without taking the lower VLA detection probability 
into account.  Thus, when all the $C_p$'s are set to 1 (\ie\ when no 
completeness corrections are included), the qualitative behavior of $\pin$, 
$\kin$, $\asymp$ and $\asymk$ is unchanged, although, as is to  be expected, 
signals are then smaller.

Systematic errors in the completenesses are not likely to be the source of
this signal.  Suppose errors of the form $C_{\rm ATCA}^\prime =C_{\rm ATCA}
(1 + \delta C_{\rm ATCA})$ and/or $C_{\rm VLA}^\prime =C_{\rm VLA}
(1 + \delta C_{\rm VLA})$ were present, and suppose further that 
the \ohir\ stars sample
only an axisymmetric part of the disc.  It is then easy to show that 
$\delta C_{\rm VLA} - \delta C_{\rm ATCA} \simeq \pin/\pin_{-}$ which is 
approximately equal to $-17/35$ in our case.  Since the noise errors are 
random, such a large systematic error would have to be in the
values of the areal completenesses, which corresponds to roughly an 
overestimate (underestimate) by $\sim 20\%$ in the radius at any given PBR for
the ATCA (VLA).  Such a large error is impossible since the PBR of both
instruments is well determined.  A systematic effect might be introduced by
the different channel widths of the VLA and the ATCA, but the similarity of
the flux density distributions between the VLA and the ATCA side suggests
that, at most, this is a $10\%$ effect.  Furthermore, this effect acts to
under-estimate the flux density of the sources detected by the VLA, so that
$f_{min}$ would remove from our sample more VLA stars than ATCA stars, which
would make $\pin$ negative, which is the opposite sign from 
what we found.  

Further evidence that the signal of Fig. \ref{fig11} is real
comes from the dependence of the signal on $b$.  We explored the effects of 
limits in $b$, finding that most of the non-axisymmetric signal comes from 
very close to the MWG plane, with little signal at $|b| \geq 1\degrees$ (which 
divides our sample of \ohir\ stars into two roughly equal parts).  This is 
unsurprising for a disc population: at 6 kpc, $1\degrees$ corresponds to 
roughly the disc scale-height.  We found considerable fluctuations in the 
value of $\kin/\pin$ at $l_0 \geq 35\degrees$ obtained from 
$|b| \geq 1\degrees$, reinforcing the impression that the settling of 
$\kin/\pin$ in the previous experiments is not caused by systematic errors, 
\eg\ in the completenesses.

All these facts make it likely, therefore, that the signal is intrinsic to 
the sample.  Finally, it is perhaps worth drawing attention to the suggestive 
similarity between Fig. \ref{fig05} and Fig. \ref{fig11}.

Although the derived value of $\kin/\pin$ is nearly constant in the range 
$35\degrees \leq l_0 \leq 45\degrees$, we cannot completely exclude 
that extending the survey to larger $l_0$ would not lead to a different 
$\kin/\pin$.  However, the density of \ohir\ stars is decreasing rapidly at 
$l_0 = 45\degrees$: an extrapolation of the binned surface density 
suggests that their surface density will vanish at around $l_0 \simeq
60\degrees$, as is also suggested by the dynamical models of Sevenster 
\etal\ (2000).  Observationally, Le Squeren \etal\ (1992) found 
that the distribution of \ohir\ stars is concentrated inside 
$|l| \leq 70\degrees$, with only roughly $15\%$ of sources outside this 
range.  Physically, this steep drop-off is the result of a 
metallicity gradient in the MWG disc: at the lower metallicities typical 
of the outer disc, AGB stars are more likely to evolve into carbon stars.  
Thus, even if there is still non-axisymmetric structure outside 
$l = \pm 45\degrees$, it is not likely to change the derived value of 
$\kin/\pin$ for the \ohir\ population very significantly.

\subsection{Missing pointings and radial velocities}

The error introduced by the missing VLA pointings is systematic in $\kin$ 
and $\pin$, in that it always reduces the VLA (\ie\ positive) 
contribution to these terms; however, the sign of the resulting change in 
$\kin/\pin$ is not obvious.  Of 88 missing pointings, 18 are in the overlap 
region.  In general, therefore, we chose to use the ATCA pointings in the 
overlap region, to minimize the effect of missing pointings.  To estimate the 
effect of the remaining missing pointings (70 of 88), we tried to reconstruct 
the missing pointings by reflection about the mid-plane where possible (49 
pointings).  Where this was not possible (21 pointings), we have simply 
averaged the value in the 4 nearest pointings, or the subset of them which 
were not themselves missing pointings.  With these changes but all else as 
before, $\pin$, $\kin$, $\asymp$ and $\asymk$ increase somewhat, but 
$\kin/\pin$ changes to $239 \pm 30$ \kms, which is not significantly different
from our previous values.  The 
reason for this robustness to missing pointings can be understood by comparing 
the number of pointings and of \ohir\ stars.  Our sample contains 73 \ohir\ 
stars detected in 840 good VLA pointings.  With these probabilities, we 
estimate that $\ltsim 70 \times 0.09 \sim 6$ \ohir\ stars have been missed 
because of missing VLA pointings not in the ATCA overlap region.

Due to observational constraints, we considered only \ohir\ stars with 
$|v_r| \leq 280$ \kms.  We experimented with reducing the maximum $|v_r|$ 
further to test how our limit might have affected our value of $\kin/\pin$.  
Radial velocities this large are expected to occur only in the bulge region, 
and indeed, the biggest relative changes in $\kin$ were at small $l_0$, but 
these propagated to small changes in $\kin$ at large $l_0$.  It is only at 
$|v_r| \leq 250$ \kms\ that stars begin to leave the sample, and $\kin/\pin$ 
decreases slowly for smaller maximum $|v_r|$, reaching $229 \pm 40$ \kms\ at 
$|v_r| \leq 200$ \kms, which is still smaller than $1\sigma$.  We therefore 
conclude that it is unlikely that our modest cuts in $|v_r|$ has resulted in 
a large systematic error in $\kin/\pin$.

\subsection{Radial motion of the LSR}
\label{lsrradmot}

A non-zero radial velocity of the LSR, $u_{\rm LSR}$, cannot be excluded.  For 
our \ohir\ star sample, $\smt/\pin\simeq 18$.  Thus even a quite small 
$u_{\rm LSR}$ will drastically alter the value of $\DV$: if $u_{\rm LSR} 
\simeq 14$ \kms, then $\DV$ changes by $100\%$.  An accurate measurement of 
$u_{\rm LSR}$ is therefore needed.  

Averaging the 197 \ohir\ stars at $-10\degrees \leq l \leq 10\degrees$ (\ie\ 
the ATCA-bulge survey), we estimated $u_{\rm LSR} = +2.7 \pm 6.8$ 
\kms\ (the plus sign indicating that the motion is away from the Galactic 
centre).  Similarly, Kuijken \& Tremaine (1994) used a variety of tracers 
(\ohir\ stars, globular clusters, high velocity stars, planetary nebulae, 
\etc) to obtain $u_{\rm LSR} = -1 \pm 9$ \kms.  Perhaps the best 
constraint on $u_{\rm LSR}$ is that of Radhakrishnan \& Sarma (1980), who 
showed that the main component in the HI absorption spectrum of Sgr A has a 
mean line-of-sight velocity $\overline{v_R} = -0.23 \pm 0.06$ \kms.  Since 
this absorbing material is presumably outside the inner region of the MWG and 
within 6-7 kpc from the Sun, this means that a wide band of material is 
moving at a common radial velocity.  While it is possible to construct models 
in which $d {\overline v_R}/d R$ is zero but with non-zero $u_{\rm LSR}$ 
(\eg\ Blitz \& Spergel 1991), the most natural explanation is that 
$u_{\rm LSR}$ is zero to this level; if $u_{\rm LSR} = -0.23$ \kms, 
this changes $\DV$ by $-4$ \kms.

\subsection{Sampling experiments and error estimates}

\begin{figure}
\begin{center}
\leavevmode{\psfig{figure=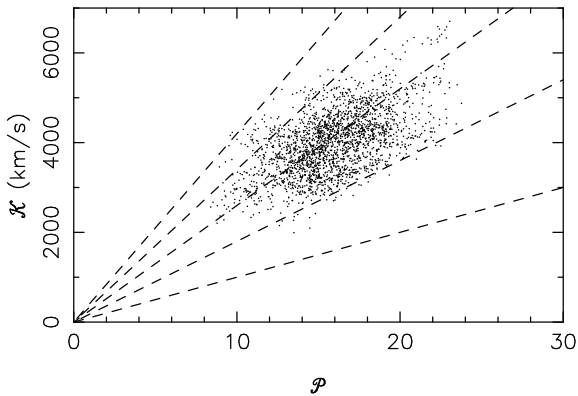,width=0.9\hsize,clip=,angle=-90}}
\end{center}
\caption{The distribution of values of $\kin$ and $\pin$ in re-sampling 
experiments with 220, 230 and 240 \ohir\ stars.  The list of stars is 
first sorted by $V_{\rm e}$, then consecutive stars are sampled in the 
range $10 \leq V_{\rm e} \leq 15$ \kms.  The dashed lines indicate 
$\om = $ 40, 50, 60, 70 and 80 \kmsk.}
\label{fig14}
\end{figure}

We now estimate the errors arising from sampling, which should be the largest
source of errors for our small sample.  We do this by re-sampling the data; 
however, since arbitrary 
resamplings will increase the noise, we first sorted the \ohir\ stars by 
$V_{\rm e}$, then took sub-samples of 220, 230 and 240 consecutive stars.  At 
each sub-sample size, we generated 1000 different reshufflings of the 
$V_{\rm e}$'s (since we always used $10 \leq V_{\rm e} \leq 15$ \kms, the 
number of combinations of \ohir\ stars, and therefore the number of different 
values of $\kin$ versus $\pin$ is smaller than 1000).  The results are 
presented in Fig. \ref{fig14}, which shows clearly that there is quite a 
fair amount of scatter in the values of $\kin$ and $\pin$.  The median value 
of $\DV = 252 \pm 41$ \kms; this corresponds to $\om = 59 \pm 5$ \kmsk.  
Therefore, as an estimate of the sampling noise, we adopt 5 \kmsk.

The error estimate on $\DV$ from sampling is consistent with the results we
found in \S\ref{tests}, and with the estimate based on variations in $l_0$.
We further test this error estimate by taking a sub-sample of PBR = 0.6, 
which gives about 180 stars.  Then the average $\DV = 175 \pm 23$ \kms, which
is within $2\sigma$ of the result for the full sample.  Since this is an 
even smaller sample than our main sample, it should be even noisier than 
our full sample.  We conclude that our sampling error error estimate is 
reasonable.  (Higher PBR levels give even smaller samples and require larger 
areal corrections (see Table \ref{tbcpa}), which make them too susceptible to 
noise.  We have therefore not attempted to use PBR levels above 0.6.)

We have experimented with other sub-samples; in all cases we found that the
results were not statistically different from the results we presented above.
However, we found that the average $\om$ tends to systematically increase as
the upper cutoff on $V_{\rm e}$ is decreased.  Thus, for example, we find that
for $10 \leq V_{\rm e} \leq 14.5$ \kms\ (212 stars), the average $\om = 64$ 
\kmsk\ while for $10 \leq V_{\rm e} \leq 14.0$ \kms\ (181 stars), the
average $\om = 69$ \kmsk.  The same trend was seen when we increased the upper
$V_{\rm e}$ to 16 \kms, although in that case we are less certain that the 
population is relaxed.  While all these results are consistent with 59 \kmsk\ 
within $2\sigma$, the systematic behavior suggests that there is a systematic
error, perhaps due to the population not being fully relaxed.  We take as an
estimate of this error 10 \kmsk.

\section{Discussion and Conclusions}
\label{discons}

The results of our study of the \ohir\ stars in the ATCA/VLA OH 1612 MHz 
survey give us a pattern speed of $\om = 59 \pm 5 \pm 10$ \kmsk\ (internal
and estimated systematic errors respectively) for $u_{\rm LSR} =0$, 
$R_0 = 8$ kpc and $V_{\rm LSR} = 220$ \kms.  If the rotation curve stays flat 
between the region 
of the non-axisymmetry and the Sun, then the corotation radius is 
$3.7 \pm 0.3_{-0.5}^{+0.8}$ kpc, while the main non-axisymmetric feature in 
question appears to be centered at $l \sim 30\degrees$, \ie\ $\sim$ 4.0 kpc 
if it is seen tangentially.  If this is the full extent of this 
non-axisymmetric feature, then $\vpd = 0.9 \pm 0.1_{-0.1}^{+0.2}$, which is 
fast in the usual definition.  However it is possible that the feature is 
not being viewed tangentially and extends somewhat further out, in which 
case $\vpd$ would be smaller.

Which feature in the MWG disc can be responsible for the signal we have 
measured?  While the TW method has the great advantage of being model 
independent, the result is that, in the absence of distance
information, it is hard to identify the feature responsible for the pattern 
speed measured.  For such a fast feature, the MWG bar would seem to be the 
natural explanation.  However, the longitude at which the non-axisymmetric 
signal peaks is 
somewhat larger than what present models of the MWG would have (\eg\ Gerhard 
2001, but see Hammersley \etal\ 2000, who have recently claimed evidence of 
a second, larger bar in the MWG).  

The small latitude of the main part of the signal implies a disc source,
while the large longitude suggests the signal arises mostly from spirals, 
particularly at the tangent point of the Scutum spiral arm.  
The high value of $\om$ and the 
relatively small $\vpd$ may hint at a coupling to the 
bar; perhaps it is even an inner ring, rather than a spiral arm.  Such rings 
are often found in external barred galaxies (Buta 1995), and are elongated 
along the bar, so that they co-rotate with the bars which they contain.  Such 
a ring has been postulated in the MWG (\eg\ Sevenster \& Kalnajs 2001).

The pattern speed measured must, in fact, be a density and asymmetry weighted
average of all the non-axisymmetric structure in the survey region, which is
known to include a bar, various spirals and perhaps a ring.  Note that there
is no evidence of substantial cancellation of $\pin$ in Fig. \ref{fig11}.  
For various sub-samples, we find some marginal evidence for multiple pattern 
speeds: when we divided our full
sample into bright (flux density $ > 0.854$ Jy) and faint (flux density 
$ < 0.854$ Jy) sub-samples, we found $\om = 49 \pm 6$ \kmsk\ and 
$\om = 71 \pm 14$ \kmsk\ respectively, together with some evidence of signals
peaking at different $l_0$.  Since the bright (faint) sub-sample is, on
average, expected to be closer to (further from) the Sun and further from
(closer to) the Galactic centre, the lower (higher) pattern speed is not 
unexpected.  However, these sub-samples are even smaller than our already 
small full sample, and the differences are not statistically significant.  
Multiple pattern speeds can be better studied with larger samples of 
tracers.

\subsection{Future prospects}
\label{future}

In the future, it will be possible to improve on our measurement in a variety
of ways.  Our main limitation was the small sample of 
objects; larger samples will provide one means of improving on our 
measurement.  At present, the ATCA/VLA OH 1612 MHz survey is the only large 
scale complete survey of a population satisfying the continuity equation.  
Kuijken \& Tremaine (1991) applied the TW method to the Galactic HI, which 
covers the entire disc, but this is unlikely to satisfy continuity, 
particularly in the presence of a bar.  However, systematic surveys of other 
populations are now in progress (\eg\ planetary nebulae Beaulieu \etal\ 2000; 
SiO masers Deguchi \etal\ 2001) or planned (Honma \etal\ 2000); these may be 
used for similar measurements.  

Accurate distance information will lead to an important refinement in the 
method.  Future astrometric satellites, such as ESA's {\it GAIA}, will 
measure radial velocities together with distance.  Not only will {\it GAIA} 
provide samples many orders larger than the ATCA/VLA OH 1612 MHz survey, but 
it will also be possible to drop the assumption of one pattern speed.  Then, 
if the pattern speed is some function of Galacto-centric radius only, we can 
write
\begin{eqnarray}
& \int \om(\rho) \mu(\rho,\psi) \cos\psi \rho^2 f(R) d\psi d\rho & 
\nonumber \\ & = \int \mu(\rho,\psi) \rho v_R(\rho,\psi) f(R) d\psi d\rho, &
\end{eqnarray}
where $(\rho,\psi)$ are Galacto-centric polar coordinates, $R$ is the usual
distance from the Sun, and for simplicity we have ignored both the third 
dimension and LSR motion.  This is a Fredholm integral equation of 
the first kind for the unknown function $\om(\rho)$; techniques for the
solution of such equations, including numerical methods, are well established.
The main difficulty with a solution of this equation probably arises in
regularising the solution at helio-centric radii $R$ at which the integrals 
are small or vanish, as 
must happen for a spiral, but a detailed discussion of these issues is
beyond the scope of this paper.  Such data will make it possible to test 
theories of spiral structure in greater detail than has been possible to
now.

\subsection{Conclusions}

We derived a 3D version of the Tremaine-Weinberg method for the Milky Way
Galaxy.  The method, as developed here, is based on the assumption that the 
density distribution can be expressed as in Eqn. \ref{dens}.  Thus we have
assumed one pattern speed only; if multiple pattern speeds are present, 
then an average pattern speed is measured, as discussed in 
\S\ref{complications}.  Moreover, we have assumed that any rapidly growing 
structure present has low amplitude.  We argued that this is likely to be 
the case since large amplitude features cannot sustain large growth for long.

We tested the method on simple models of bars and spirals, selecting samples 
of 500 discrete tracers, to show that it is possible to measure pattern 
speeds with average error of $17\%$, and always better than $40\%$, provided
that asymmetry signals are sufficiently large, as described in 
\S\ref{tests}.  We then 
extracted a sample of some 250 \ohir\ stars from the ATCA/VLA OH 1612 MHz 
survey and applied the method to these stars.  This gave 
$\kin/\pin = 252 \pm 41$ \kms, from which we obtain $\om = 59 \pm 5$ \kmsk, 
with a possible systematic error of perhaps 10 \kmsk, 
if $V_{\rm LSR} = 220$ \kms, $R_0 = 8$ kpc and $u_{\rm LSR} = 0$ (for other 
values of $V_{\rm LSR}$, $R_0$ or $u_{\rm LSR}$, $\om$ can be obtained from 
$\om R_0 - V_{\rm LSR} = 252 \pm 41 - 18 u_{\rm LSR}$).  The sample is 
quite small and the detection is only significant at about 
the $90\%$ level.  Future larger samples will improve this situation; we 
sketch how distance data combined with the projected position and line-of-sight
velocity data used in our study will lead to a substantial improvement in our
knowledge of the pattern speed(s) in the Milky Way Galaxy.  

\bigskip
This work was partly supported by grant \# 20-56888.99 from the Swiss National 
Science Foundation.  We thank the anonymous referee for comments which 
helped improve this paper.

\label{lastpage}

\end{document}